# Latent Class Logit Kernel Framework for Surrogate Safety: Identifying Behavioural Thresholds through Conflict Indicator Profiles

*Rulla Al-Haideri[1], Changhe Liu[2,3], Karim Ismail[2], Bilal Farooq[1], Chi Zhang[3]*
[1]Laboratory of Innovations in Transportation (LiTrans), Toronto Metropolitan University, Canada,
[2]Carleton University, Canada, [3]Chang'an University, China

**Abstract**

Crash data objectively define and characterize road safety but are rare and often unsuitable for proactive safety management. Traffic conflict indicators, such as time-to-collision (TTC), offer continuous measures of proximity to collision but require thresholds to distinguish between routine and safety-critical interactions. Extreme Value Theory (EVT) provides statistically defined thresholds, but these values do not necessarily reflect how drivers perceive and respond to escalating conflict. This study introduces a new behavioural modelling framework that identifies candidate behavioural thresholds (CBTs) by explicitly modelling how drivers adjust their movements under conflict conditions. The framework is built on a Latent Class Logit Kernel (LC-LK) model, which goes beyond the structure of a standard multinomial logit. It captures inter-class heterogeneity by distinguishing between routine and defensive driving styles, and intra-class heterogeneity through a cross-nested error structure that allows correlation between overlapping spatial alternatives (directional choices) and speed adjustments. This design produces probability curves, that describe how the likelihood of defensive manoeuvres changes with conflict indicator values. From these profiles, CBTs such as inflection points, crossovers, and tail-based thresholds can be derived. The framework is guided by four behavioural hypotheses: (*i*) drivers simultaneously express varying degrees of membership in both low-risk and high-risk behavioural classes; (*ii*) class membership shifts systematically with conflict indicator values, revealing thresholds where behaviour changes more sharply; (*iii*) this relationship often follows a logistic shape, with stable behaviour across safe conditions and rapid transitions once critical values are reached; and (*iv*) even in free-flow conditions, drivers maintain a baseline level of caution, ensuring that single-vehicle risks and inherent risk aversion are also captured. Application to naturalistic roundabout trajectories illustrates the framework. For TTC, probability curves revealed stable thresholds (inflections around 0.8–1.1s), while a modified form of TTC (MTTC2) showed instability in the full dataset, with implausibly high crossovers (*e.g.*, 34s). This divergence motivated a tentative cognitive load hypothesis: thresholds may only emerge when drivers are cognitively able to process the underlying indicator. TTC, requiring only constant-velocity projection, yielded interpretable thresholds, whereas MTTC2, requiring simultaneous evaluation of speed, acceleration, and curvature, exceeded cognitive limits under routine driving. Overall, the framework provides a structured and behaviourally grounded complement to EVT. The framework integrates inter- and intra-class modelling with diagnostic tools and stability checks, demonstrating how behavioural thresholds can be identified, validated, and meaningfully interpreted in surrogate safety analysis.

## 1. Introduction

Traffic safety research has long relied on crash data as the primary evidence base for identifying hazardous conditions and prioritizing interventions. While crashes are concrete and well-defined events, they are rare events, often requiring years of data to achieve sufficient statistical power. Crash databases are vulnerable to underreporting, inconsistencies in severity classification, and lags in availability. Consequently, decision-makers need timely and sensitive measures to diagnose risk before crashes occur. To address this gap, the field has increasingly turned to surrogate safety assessment to capture near-crash interactions that occur with greater frequency. Analysing how vehicles or other road users interact in everyday conditions allows researchers to infer the likelihood of more severe outcomes. This could bridge the gap between operational performance and long-term safety.

Traffic conflict indicators quantify the proximity of two road users in temporal or spatial terms and are sensitive to relative speed, trajectory, and spacing. Their appeal lies in their ability to provide a continuous measure of safety across thousands of interactions, rather than relying on binary crash outcomes. For example, Time-to-Collision (TTC) values approaching zero indicate increasing collision likelihood. Post-encroachment Time (PET) measures capture whether one road user clears a conflict point before another arrives. Such measures have been widely applied in naturalistic driving studies, microsimulation analyses, and trajectory-based datasets that enables safety assessments in complex environments such as intersections, roundabouts, and freeway merges. However, a persistent challenge in using conflict indicators is the determination of thresholds. A conflict indicator is only useful for safety diagnosis when researchers can reliably distinguish between "safe" and "unsafe" values. In practice, thresholds have been set in diverse ways including engineering judgement (*e.g.*, defining TTC < 1.5 seconds as critical), empirical percentiles of observed distributions, or using statistical approaches. Thresholds are not trivial choices, they strongly influence the number and type of events considered safety-critical, and by extension, the estimated safety performance of a site or the evaluation of a countermeasure. For example, a conservative TTC threshold will classify a large number of interactions as unsafe which could potentially overstate risk. However, a more lenient threshold may overlook meaningful precursors to crashes. Thus, threshold selection remains a bottleneck in conflict-based safety analysis.

Extreme Value Theory (EVT) has provided powerful tools to formalize this problem (Botvinick et al., 2001; Shenhav et al., 2013); Ali et al., 2022; Zheng et al., 2019). EVT offers methods for modelling the tails of distributions and estimating the probability of rare but extreme events. In traffic safety, EVT is employed to quantify the likelihood of severe conflicts by focusing on the extreme values of conflict indicators. The Peak-Over-Threshold (POT) method, in particular, has become influential. The exceedances beyond a chosen threshold are modelled using the Generalized Pareto Distribution (GPD). POT allows researchers to define a threshold that separates "safety-relevant" from "non safety-relevant" events. Because EVT focuses on the tail of the distribution, this threshold is typically set at very small indicator values, and all events above this cut-off are discarded as routine and non-critical. This approach ensures statistical rigour, reduces subjectivity in threshold choice, and enables extrapolation to rare conflict scenarios. Its adoption in traffic safety research has marked a major step forward, replacing ad-hoc thresholds with defensible, data-driven ones.

Despite its strengths, the POT approach has important limitations in behavioural contexts. Thresholds are typically defined as statistical cut-offs, without explicit consideration of how drivers perceive or respond to risk. For instance, a TTC of 2 seconds may be statistically safe on average, but drivers may still engage in early braking or evasive manoeuvres before this value is reached. Conversely, some drivers may delay reacting until the indicator approaches dangerously low values. By treating thresholds as purely statistical artefacts, POT overlooks the heterogeneity in driver decision-making and the possibility that behavioural transitions occur at points that do not align with statistical thresholds. POT is designed to capture rare exceedances and therefore provides limited insight into the more frequent, moderate-risk interactions that are highly relevant for proactive safety management. In trajectory datasets, decisions are shaped by temporal and spatial dependencies, the assumption that exceedances are independent and identically distributed is also problematic.

These limitations motivate the exploration of complementary approaches that explicitly incorporate driver behaviour into threshold determination. While EVT provides a statistical framework for defining thresholds, behavioural modelling approaches from discrete choice analysis could also serve as complementary tools for capturing decision-making mechanisms. Two notable methodological directions in discrete choice modeling provide the foundation for the approach developed in this study. Greene and Hensher (2013) introduced the Random Parameters Latent Class (RPLC) model, which captures two layers of heterogeneity. These layers include discrete differences between behavioural classes (*e.g.*, aggressive versus cautious drivers) and continuous variation within each class through random parameters. Ben-Akiva et al. (2001) developed the logit-kernel model, which uses a hybrid error structure combining multivariate normal components (to capture correlation among alternatives) with independent Gumbel errors (to maintain computational tractability).

In this study, we synthesize these approaches in a novel way. We employ latent classes to capture discrete behavioural regimes (following the spirit of Greene and Hensher (2013)'s inter-class heterogeneity). We also use Ben-Akiva's logit-kernel error structure within each class to capture correlation among similar driving maneuvers. Specifically, the latent classes distinguish between defensive and routine driving classes based on conflict severity. The kernel error structure accounts for the fact that similar maneuvers (*e.g.*, all deceleration choices or all turning movements) are correlated through shared unobserved factors. This combination of discrete behavioural regimes with structured error correlation, forms what we term the Latent Class Logit Kernel (LC-LK) model. Behavioural thresholds are then inferred from points along the conflict indicator profile where probability of class membership shift sharply, which signals transitions from low- to high-risk driving states.

The thresholds obtained from EVT are defined as statistical cut-offs rather than behavioural markers. In practice, this means that any conflict indicator value below the cut-off is automatically treated as safety-relevant, while all higher values are ignored. Such a rule may classify controlled but safe interactions as "critical," while overlooking moderate but behaviourally important responses. Tarko et al. (2009) and Tarko (2012, 2018, 2021) emphasizes that traffic events need to be carefully distinguished according to their relationship with crashes. Traffic encounters include all interactions between road users where the majority of which are routine and safe. A smaller subset of these encounters involves conflicts. These are defined as situations where the spatio-temporal proximity implies a non-zero probability of collision if movements remain unchanged (Amundsen & Hydén 1977). Even within conflicts, some events are more relevant to safety than others. Safety-relevant events are those conflicts that originate from temporary lapses in hazard perception or delayed response and that carry a potential for harmful outcomes (Glauz & Migletz, 1980); Tarko, 2021)). These are often characterised by unexpected evasive manoeuvres where consequences would likely be serious enough to be reported as crashes. In contrast, some aggressive but controlled interactions (*e.g.*, slow following or low-speed lane clearances) may be identified as conflicts according to an indicator but are less meaningful as precursors to crashes.

A threshold chosen purely from the distribution of an indicator may capture both error-driven, safety-relevant events, while simultaneously including controlled manoeuvres that do not share the same cause with crashes. Extrapolating from a mixed sample of events can be problematic, because the link between observed events and crashes depends on their etiological consistency. Based on those insights, we concur that extreme indicator values tied to failures and harmful outcomes represent a step higher in the safety hierarchy. As Tarko (2021) notes, a key challenge for EVT-based approaches is ensuring that the events used for tail modelling are indeed safety-relevant. Similar considerations have been raised in counterfactual approaches, which emphasize that valid crash precursors must reflect failures rather than deliberate but safe driving strategies (Davis et al., 2011).

From this perspective, there is scope to complement EVT with approaches that account for how drivers interpret the environment and respond to risk. Building on these insights, this paper proposes a behavioural modelling framework that identifies "behavioural" thresholds based on observed driver responses to conflicts. Specifically, we examine the potential of a LC-LK approach for identifying behavioural thresholds in traffic conflict indicators. Discrete Choice Models (DCMs), grounded in random utility theory, provide a well-established methodology for representing decision-making under uncertainty (Al-Haideri et al., 2025a; Al-Haideri et al., 2025b; Al-Haideri et al., 2025c). In driving contexts, DCMs allow researchers to model how road users select manoeuvres, such as braking, accelerating, or maintaining course, based on perceived conflict intensity and environmental cues. Extending DCMs to a latent class formulation may enable the identification of distinct behavioural states, such as low-risk (routine driving) and high-risk (defensive or evasive) classes, while also capturing unobserved heterogeneity across drivers and interactions. The contribution of this approach lies in linking class membership (CM) probabilities directly to conflict indicators. Instead of imposing predefined statistical cut-offs, thresholds in this study are identified at points where the probability of belonging to the high-risk class changes sharply, signalling a behavioural shift in driver response. This may reveal a profile of behavioural thresholds that serve as behaviourally interpretable benchmarks to statistical cut-offs. In contrast, the behavioural thresholds identified in this study indicate the points at which drivers begin to change their responses to distinguish safety-relevant interactions based on observed behaviour rather than only statistical rarity. Such thresholds could offer insights into when drivers begin to adjust their behaviour under moderate risk and when their responses accelerate as conflicts become more severe. These points emerge from observed behaviour in naturalistic datasets, rather than from rigid thresholds imposed by existing methods, but their interpretation remains context-dependent.

The objective of this study is not to replace established EVT methods, but to introduce an alternative perspective grounded in behavioural modelling. EVT offers a statistically rigorous foundation for modelling extreme events, while the LC-LK framework emphasises the mechanisms through which drivers perceive and respond to escalating conflict. Although these approaches are not integrated in the present study, they could potentially serve as complementary tools in future research, one capturing statistical rarity, the other providing behavioural interpretability. Framing them in parallel underscores the value of considering both perspectives when developing proactive safety management strategies, where the aim is not only to estimate the probability of rare and severe outcomes but also to understand how everyday interactions evolve toward hazardous situations. Building on Hydén's safety pyramid, traffic conflicts can be viewed as a continuum of events, each representing a different level of severity (Hydén, 1987). Conflict indicators capture specific facets of this continuum, and in this paper, we focus on a univariate representation. We hypothesize that driver decision-making varies systematically with conflict severity, and such variation can be represented through latent CM probabilities. Modelling the relationship between a conflict indicator and behavioural class probabilities allows the framework to identify thresholds that are both data-driven and behaviourally interpretable.

We therefore examine three guiding questions: (*i*) to what extent can the relationship between a conflict indicator and latent driver behaviour classes be used to identify candidate thresholds that might separate different decision-making regimes; (*ii*) do these thresholds appear to reflect behavioural shifts rather than artefacts of statistical specification; and (*iii*) how might the inclusion of the full dataset, which contains both conflict events (where the indicator is defined) and non-conflict events (such as free-flow conditions), influence threshold estimation compared with conflict-only subsets. The methodological contributions of this work are to:

1. propose a LC-LK model that links CM probabilities to conflict indicators as a way to explore behavioural thresholds,
2. suggest procedures to extract candidate behavioural thresholds (CBTs) that may reflect systematic shifts in driver response as conflict severity increases, and
3. outline a validation process that could help assess the robustness and behavioural interpretability of identified thresholds.

The paper is organized as follows: Section 2 outlines the proposed framework, including the hypotheses, model specification and the procedures for identifying and validating behavioural thresholds. Section 3 presents an application to the proposed framework and discusses the results, and Section 4 concludes with key findings and future research directions.

## 2. Proposed Framework

This section presents the behavioural threshold identification framework, starting with the theoretical hypotheses, followed by the model structure and specification, estimation procedures, and finally the methods for identifying and validating behavioural thresholds.

### 2.1. Behavioural Thresholds Hypotheses

The aim of this paper is to identify behavioural thresholds for a traffic conflict indicator, defined here as values at which drivers exhibit noticeable shifts in their responses to a situation. To investigate this, we adopt a LC-LK framework. In this framework, the driver's behavioural pattern or class is represented through CM. Each class reflects a distinct mode of decision-making, such as routine (low-risk) driving or defensive (high-risk) behaviour. For parsimony, we assume two latent classes in this study; however, exploring richer class structures with more behavioural segments is left for future work. The CM varies as a function of the conflict indicator being studied. As the indicator changes, the probability of belonging to one class or the other may shift, which provides a way to trace how drivers perceive risk and adjust their responses.

In this study, CM is specified using a single conflict indicator in a univariate formulation. This makes it possible to link changes in behavioural class probabilities directly to one safety-related measure. Estimating these probabilities across the indicator's range can reveal points where behavioural responses change more sharply. We interpret these points as behavioural thresholds, offering behaviourally meaningful values to statistical cut-offs defined in EVT. For a given indicator, we define not just a single threshold but a profile of thresholds, where each point in the profile marks a potential behavioural shift. To assess whether these shifts are meaningful, we also apply diagnostic checks to evaluate the evidence for behavioural change around each candidate threshold. Based on this framework, we put forward the following hypotheses:

- Hypothesis 1: At any given moment, a driver can simultaneously belong to two latent behavioural classes with varying degrees of membership: a low-risk class, characterised by routine or comfortable driving, or a high-risk class, characterised by defensive or evasive responses.

- Hypothesis 2: The probability of belonging to the low-risk or high-risk class changes systematically with the conflict indicator. At specific values of the indicator, this probability shifts more noticeably, and these points can be interpreted as behavioural thresholds.

- Hypothesis 3: The relationship between the probability of being in the high-risk class and the conflict indicator may follow a logistic shape. Drivers remain in the low-risk class across a wide range of moderately safe conditions, but once the indicator reaches a critical level, the probability of shifting to the high-risk class increases sharply.

- Hypothesis 4: Even in free-flow conditions or when driving alone, drivers exhibit a baseline probability of caution that reflects their inherent risk avoidance tendencies. Including these events in the analysis ensures that the CM model captures this underlying caution. Drivers maintain a minimum level of caution to avoid risks, including single-vehicle crashes, even without the presence of an immediate conflict.

These hypotheses establish the foundation for how behavioural thresholds are conceptualised in this study. Hypothesis 1 emphasises that drivers are not rigidly assigned to a single behavioural class but instead express a degree of membership in both low-risk and high-risk classes at any moment. Hypotheses 2 and 3 extend this by proposing that these memberships shift systematically with changes in the conflict indicator. In this case, the relationship takes a logistic shape where responses remain stably high at small indicator values, decline more noticeably across a mid-range of values, and then stabilise again at lower risk levels once the indicator becomes large. These hypotheses highlight that behavioural thresholds emerge from the interaction between gradual changes in perceived risk and sharp transitions in decision-making. Hypothesis 4 broadens the scope by recognising that behavioural thresholds cannot be inferred solely from encounters where the indicator is computable. Including the all driving contexts, which also contains non-collision-course and free-flow situations, allows the analysis to capture the role of baseline caution and single-vehicle risk avoidance in shaping driver behaviour. In practice, this baseline probability is represented in the CM specification through an alternative-specific constant (ASC), which is included alongside the conflict indicator and will be explained in Section 2.3.1.

## 2.2. Model Structure

Extensions of discrete choice models relax two main limitations of the standard MNL: the IID error assumption and the assumption of homogeneous unobserved preferences. Following Bhat and Guo, (2004), the IID assumption can be relaxed by allowing errors to be (a) identically distributed but correlated, (b) independent but non-identical (heteroskedastic), or (c) both correlated and non-identical. In practice, preference heterogeneity across decision-makers is modeled by random coefficients (Bhat & Guo, 2004) or latent classes (Greene & Hensher, 2003). While departures from IID errors are handled with structured error terms (e.g., GEV nests (Train, 2009) or shared-error components such as a logit-kernel (Ben-Akiva et al., 2001).

Our LC-LK model combines these ideas. At each decision point, drivers are a mixture of latent classes (inter-driver heterogeneity). Within each class, we add shared normal errors with group-specific scales, where "groups/nests" refer to the speed rings and turn cones defined below. The choice set consists of a 3×3 spatial grid representing nine possible driver maneuvers. It includes three speed rings (decelerate, maintain speed, accelerate) crossed with three turn cones (left, straight, right), yielding nine alternatives. An alternative belongs to exactly one speed ring and one turn cone. Within a class, all alternatives in the same speed ring share a normal error with that ring's scale, and all alternatives in the same turn cone share a normal error with that cone's scale. Two alternatives are therefore correlated if they share a ring or a cone. This setup captures correlation in a behaviourally transparent way and permits group-level heteroskedasticity through the ring and cone specific scales.

### 2.2.1. Logit-Kernel vs. Standard Cross-Nested Logit

We choose the logit-kernel structure over cross-nested logit (CNL) because it better matches our application by capturing both within and across group correlation while allowing flexible heteroskedasticity through group specific error scales. In logit-kernel, correlation occurs whenever alternatives share the same normal error component, with scales that can differ by group, creating heteroskedasticity. This flexibility is crucial for spatially related driver choices where similar maneuvers (*e.g.*, all deceleration alternatives) exhibit correlation. We explain why we use a logit-kernel model instead of a standard cross-nested logit (CNL) to capture correlation among alternatives. We compare both model structures based on how their errors are built (errors and scales), how they generate correlation and substitution, how they handle heteroskedasticity, and how identification and normalisation are set.

The two model structures build randomness differently. In the logit-kernel model, each utility combines a systematic term, an IID Gumbel error, and one or more shared normal errors whose scales can differ by group. Choice probabilities are obtained by averaging standard logit probabilities over the distribution of the shared normal errors. In CNL, all randomness comes from a Generalized Extreme Value (GEV) structure defined by nesting coefficients and allocation parameters. Here, probabilities are closed-form, and no separate normal errors are added.

We differentiate between correlation and substitution. In logit-kernel, correlation happen whenever alternatives include the same shared normal error. If the same shared error is reused across groups, correlation appears within and across groups. Because the scales on these shared errors can differ by group, error variances differ by group (heteroskedasticity). These shared errors also concentrate substitution. When one alternative becomes less attractive, probability tends to shift first to options that share the same shared error. The key uniqueness in this structure is its is flexibility, we can obtain create both within-group and across-group correlation and set group-level heteroskedasticity directly by the shared-error scales.

In CNL, alternative-level errors are Gumbel only. Correlation and substitution come exclusively from the nest structure through the nesting coefficients and allocation parameters. Therefore, substitution is stronger within the same nests, and there is no across-nest correlation unless shared membership is explicitly assigned. Heteroskedasticity is tied to the nests, not freely set per alternative. The key uniqueness of this structure is that correlation exists only where membership overlaps, and its strength is governed by the nesting design rather than added shared errors.

Identification and normalisation also differ. In logit-kernel, we follow standard practice by normalising the IID logit error scale to one. We then normalize one normal-error scale, so all other scales are interpreted relative to it (we use the deceleration ring and left-turn cone as references in this paper). In a purely heteroskedastic version with independent normal errors per alternative, a practical choice is to set the minimum-variance group to zero (further details can be found in Ben-Akiva et al., 2001). This yields a unique parameterisation and nests MNL when all normal-error scales are zero. In CNL, identification follows the usual GEV rules. The nesting coefficients are constrained to (0,1], allocation parameters are non-negative and normalised per alternative (*i.e.*, sums up to one) (Train, 2009).

We choose the logit-kernel structure because it is less restrictive and better matches our application. It captures both within- and across-group correlation while allowing flexible heteroskedasticity. Those features could be crucial for spatially related driver choices.

### 2.3. Model Specification

Formally, the probability that driver $n$ selects alternative $i$ at time $t$ is obtained by aggregating the class-specific choice probabilities, weighted by the probability of the individual belonging to each class, abbreviated by $P_{nit}$, is described as (Greene & Hensher, 2003):

$$P_{nit}(\theta) = \sum_{s=1}^{S} P_{nts}(\theta) \cdot P_{nit}|s\,(\theta)$$

where, $P_{nt(s)}(\theta)$ is the probability that driver $n$ belongs to class $s$ at time $t$ (CM probability) obtained in closed form using the standard logit function. This captures inter-driver heterogeneity. The $P_{nit}|s(\theta)$ is the probability that individual $n$ selects alternative $i$ at time $t$, conditional on being in class $s$, which has no closed form and must be approximated using simulation. This captures intra-driver heterogeneity. Here, the parameter vector ($\theta$) represents two distinct set of parameters:

- CM parameters ($\beta_{CI}$, $\beta_{ASC}$), which govern inter-driver heterogeneity and are estimated in closed form because CM probabilities have a standard logit structure.
- Conditional choice (CC) parameters, which govern intra-driver heterogeneity. These include the variance parameters of the structured error components (*e.g.*, across speed rings and turning cones) and any coefficients in the systematic utility $V_{nis}$. Because these enter through normally distributed error terms, their contribution to the likelihood requires simulation.

### 2.3.1. Inter-Class Heterogeneity Captured by Class Membership

We assume two latent classes to capture systematic behavioural differences. The high-risk class reflects stronger adjustments under conflicts (defensive/evasive tendencies), while the low-risk class reflects routine driving. The probability that driver $n$ belongs to class the high risk class ($s = HR$) is given by:

$$P_{nt(HR)} = \frac{e^{M_{nt}^{HR}}}{1 + e^{M_{nt}^{HR}}}$$

where $M_{nt}^{HR}$ represents the systematic portion of utility for the high risk class. The utility of the low risk class ($s = LR$) is normalized to zero ($M_{nt}^{LR} = 0$) for model identification purposes. The specification of $M_{nt}^{HR}$ links CM to the presence and severity of a conflict indicator, described by:

$$M_{nt}^{HR} = \begin{cases} \beta_{CI} \dfrac{1}{CI_{nkt}^{min}} + \beta_{ASC} & \text{if } CI_{nkt}^{min} > 0 \\ \beta_{ASC} & \text{if } CI_{nkt}^{min} \text{ is undefined} \end{cases}$$

where:

- $CI_{nkt}^{min}$ is the minimum value of the respective conflict indicator being analysed (*i.e.,* TTC or MTTC2) between driver $i$ and any interacting vehicle $k$ at time $t$,
- $\beta_{ASC}$ is a constant capturing the baseline propensity to belong to the high-risk class, and
- the coefficients $\beta_{CI}$ and $\beta_{ASC}$ are unknown and to be estimated.

When the conflict indicator is defined (for example, when two vehicles are on a potential collision course and a TTC or MTTC2 value can be calculated), the high-risk utility includes both the effect of the indicator ($\beta_{CI}$) and the constant term ($\beta_{ASC}$). In contrast, when the conflict indicator is not defined (for example, in free-flow conditions, parallel movements, or situations without a collision course), the high-risk utility reduces to the baseline term ($\beta_{ASC}$) only. This constant plays a critical role by ensuring that the probability of being classified as high-risk never becomes zero because no conflict indicator is observed. In other words, even in the absence of measurable vehicle–vehicle interactions, drivers maintain a baseline probability of exhibiting high-risk behaviour. This reflects the idea that a minimum level of caution is always present, capturing behaviours such as avoiding single-vehicle collisions, losing control, or making abrupt adjustments in seemingly safe conditions. This specification guarantees that the framework accounts not only for conflict-driven risk but also for inherent baseline risk present in routine or non-interactive driving. In addition, we acknowledge that this specification is univariate, since it considers a single conflict indicator and two latent classes. We recognise, however, that in reality more than two latent classes may exist and that multiple indicators could be specified jointly (bivariate or multivariate formulations). These richer extensions are reserved for future work. We also note that in the utilized dataset we did not have cases where the conflict indicators were exactly equal to 0, but if such cases were present, a specification of $(1/(1 + CI_{nkt}^{min}))$ could be adopted to maximise the utility when the indicator equals zero.

### 2.3.2. Intra-Class Heterogeneity Captured by Conditional Choice Probability

Conditional on CM, drivers may still exhibit individual variation in their decision-making. This within-class heterogeneity is captured through random error components that are structured to allow correlation among behaviourally similar alternatives. The probability that driver $n$ selects alternative $i$, conditional on being in class $s$ is expressed as:

$$P_{nit}|s = \int \frac{e^{V_{nit}^s + \eta_{nti}}}{\sum_j^J e^{V_{njt}^s + \eta_{njt}}} \phi(\eta_n) d\eta_n$$

where $V_{nit}^s$ is the systematic utility component, and $\eta_n = (\eta_{nt1}, \dots, \eta_{ntJ})$ is a vector of shared normal errors with a multivariate normal density $\phi(.)$ that indues correlation across alternatives.

The analysis relies on road user trajectory data, without requiring collision records, and is intended as an exploratory step toward incorporating behavioural interpretation into conflict-based safety analysis. At each time step, the model represents the driver's next movement as a discrete choice among a finite set of plausible options. Rather than treating the vehicle's trajectory as continuous and unconstrained, the approach assumes that drivers face a limited menu of manoeuvres, each reflecting a realistic combination of speed change and steering direction. In this study, the choice set consists of nine alternatives, where each alternative represents a possible future position in space that a driver could occupy in the next second (Al-Haideri et al., 2025b). Within the DCM framework, the probability of selecting each alternative is modelled using a utility function composed of a systematic component and a random error term. The systematic component captures measurable factors such as distance from the intended path, turning angle, or change in speed, and is evaluated separately for each alternative or future position. These utilities, conditional on latent CM, define the likelihood of a driver selecting a particular manoeuvre at a given moment.

We emphasize a key distinction between this study and the related work by Al-Haideri et al., (2025b), which employed a similar modelling framework but introduced an additional behavioural dimension: the probability of a driver being in a defensive state. CM probability was linked to variables representing a driver's general situational awareness or perceived environmental risk, such as traffic volume, density, or sustained closing patterns. When a conflict indicator was included at this level, it reflected an abstract or anticipatory assessment of risk rather than an immediate threat. By contrast, CC probabilities were driven by temporally or spatially precise indicators that captured real-time collision threats, including relative speed, heading, or proximity to surrounding vehicles. These indicators influenced manoeuvre selection only when drivers were classified as defensive, thereby shaping operational decisions such as braking or evasive steering.

Figure 1 illustrates the spatial configuration of this 3×3 grid of alternatives. The grid combines three levels of forward progression (based on observed acceleration patterns) with three levels of steering deviation (left, direction maintenance, right). The concentric arcs capture different forward displacements tied to the driver's current speed $v_{nt}$ and time step $\Delta t$, capturing variations in forward progression. The innermost arc reflects a deceleration behaviour ($1.04v_{nt}\Delta t$), the middle arc corresponds to speed maintenance behaviour ($1.07v_{nt}\Delta t$), and the outermost arc denotes an acceleration behaviour ($2.14v_{nt}\Delta t$). These boundaries are obtained based on percentile distributions of the observed change in speed and direction for a driver at the next time step in the utilized data in this paper. Detailed steps are presented in (Al-Haideri et al., 2025a). The arcs are then intersected with angular divisions to represent steering or turning behaviour. The left (cells 1, 4, 7) and right (cells 3, 6, 9) represent increasing deviations. The central (cells 2, 5, 8) represent direction maintenance. This structure ensures that the set of nine alternatives spans conservative to aggressive accelerations and mild to sharp turns, allowing the model to approximate the spectrum of manoeuvres that drivers use when responding to potential conflicts.

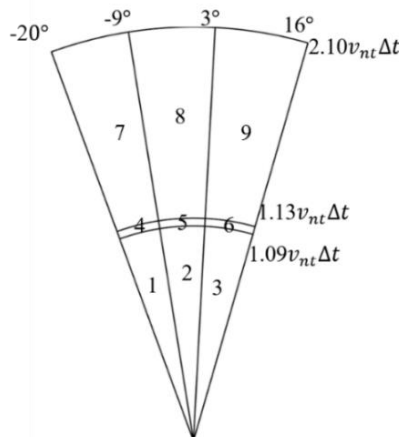

*Figure 1 Spatial choice set grid.*

To ensure that $\eta_{ni}$ reflects behavioural similarity across alternatives, we decompose it into structured components. As shown in Figure 1, each alternative belongs simultaneously to one of three speed rings (decelerate, maintain speed, accelerate) and one of three turning cones (left, straight, right). The intuition is that drivers rarely view alternatives in isolation. We assume that they consider groups of alternatives that are similar in speed adjustment and directional choice. For example, a driver inclined to decelerate is more likely to consider all deceleration options (1–2–3) in comparison to acceleration options (7–8–9). Likewise, a driver inclined to turn left is more likely to weigh left-turn options (1–4–7) against other turning directions. To capture this behavioural correlation, the total utility of alternative $i$ perceived by driver $n$ in class $s$ is specified as:

$$U_{int}|s = V_{nit}^s + \underbrace{\eta_{nt}^{Speed(a)}}_{\text{shared within speed rings}} + \underbrace{\eta_{nt}^{Turn(b)}}_{\text{shared within turn cones}} + \varepsilon_{nit}$$

where $a \in \{decel, keep, acc\}$, $b \in \{left, straight, right\}$, $\varepsilon_{nit}$ is IID Gumbel. Thus, alternatives (1–2–3), (4–5–6), (7–8–9) share a speed error. Also, alternatives (1–4–7), (2–5–8), (3–6–9) share a turn error. Each shared normal error has a group-specific scale, which drives both correlations (within and across groups) and heteroskedasticity (group-level variances). We fix one speed ring normal error scale (deceleration group), and one turn cone normal error scale as references (left-turn group). All other group scales are estimated relative to these references. Only variances (scales) of the structured errors are meaningful here, systematic differences across alternatives are already handled by $V_{nit}^s$.

### 2.4. Estimation Procedure

Because the conditional choice probability integral has no closed form, it is approximated using simulation with $R$ random draws. In this study, we set $R = 100$. Following Bhat & Guo, (2004), we generate these draws using a scrambled Halton sequence rather than purely pseudo-random numbers to improve efficiency and stability, the CC probability is expressed as:

$$P_{nit}|s \approx \frac{1}{R}\sum_{r=1}^{R} \frac{e^{V_{nits}+\eta_{nits}^r}}{\sum_j^J e^{V_{njts}+\eta_{njts}^r}}$$

The full simulated log-likelihood (SLL) function is then constructed by summing across all drivers $n$, time instant $t$, and chosen alternatives $i$ is represented by:

$$SLL(\theta) = \sum_{n=1}^{N}\sum_{t=1}^{T}\sum_{i=1}^{J} y_{nit} \cdot \ln\left(\sum_{s=1}^{S} P_{nits}(\theta) \cdot P_{nit}|s(\theta)\right)$$

where $y_{nit}$ indicates the observed choice.

### 2.5. Behavioural Thresholds Identification

#### 2.5.1. Threshold Identification Process

We hypothesize that a driver can simultaneously belong to the two latent classes: low-risk and high-risk with a varying degree at any moment in time. The low-risk class represents routine, goal-directed behaviour with minimal sensitivity to conflict. The high-risk class reflects heightened responsiveness to conflict cues and an increased likelihood of evasive or defensive manoeuvres. The model consists of two components. The first is the CM probability, which estimates the probability that a given driver trajectory belongs to either the low- or high-risk class. A conflict indicator is specified as the key explanatory variable in the CM probability. This enables the model to relate perceived conflict intensity to the likelihood of CM. The functional relationship between the conflict indicator and CM probability is then examined to detect the profile of behavioural thresholds. Statistical tests are employed in subsequent steps to assess whether the transition across this threshold reflects a significant behavioural shift. The second component of the model is the CC probability, which defines the likelihood of selecting a particular alternative from a discrete choice set, conditional on latent CM. Specifically, within each class, drivers are assumed to choose among a set of potential future positions according to a class-specific utility function. The CC probabilities include structured error components ($\eta$ terms) that capture correlation among behaviourally similar manoeuvres, as described in Section 2.3. The utility functions represent the systematic (*i.e.*, observable) part of decision-making and differ in structure between the two classes to reflect their underlying behavioural priorities. In the high-risk class, the utility function incorporates variables that indicate increased conflict response and urgency, such as rapid deceleration, lateral movements near other vehicles, or large deviations from the reference path. These terms capture risk-averse or evasive tendencies. In contrast, the low-risk class prioritises goal-oriented driving under normal conditions, with utility variables reflecting stable trajectory following, consistent speed, minimal turning, and path adherence. The distinct utility functions in each class ensure that the CC probabilities reflect the different behavioural mechanisms governing driver decisions under varying levels of perceived conflict.

The choice of functional form for the conflict indicator in the CM probability is critical. This function governs how the severity of the conflict is perceived and modelled. This functional form shapes the gradient by which the indicator influences the probability of assignment to the low- or high-risk behavioural class. Since conflict indicators such as TTC often span wide numeric ranges, their raw scale may not directly correspond to behavioural perception. For example, a linear specification may underrepresent the urgency of shorter TTC values, which are behaviourally more critical, where evasive action is most likely, while exaggerating the importance of longer, non-critical values. Beyond these basic transformations, scaling the indicator by raising it to a power can further modulate its sensitivity. For example, using $1/TTC^2$ instead of $1/TTC$ disproportionately increases the influence of very short TTC values, potentially capturing the steep escalation in perceived urgency as a collision becomes imminent. The exponent in such transformations can either be selected based on behavioural reasoning, for instance, to reflect a nonlinear escalation in driver response, or empirically estimated from the data as an additional parameter. However, overly aggressive scaling may distort the behavioural gradient or lead to estimation instability, so the choice or estimation of the exponent must be approached with care. Normalising the indicator, such as by dividing TTC by a reference value like a driver's perception-reaction time (*e.g.*, TTC/2.5 s), adjusts the scale to reflect individual or context-specific sensitivity. This approach enables comparisons across drivers or situations with different reaction capabilities. Normalisation can also make the resulting transformation dimensionless, improving interpretability and comparability. For instance, a TTC of 2.5s divided by a reaction time of 2.5s yields a value of 1, which could serve as a behavioural reference point for imminent action. Each candidate functional form is evaluated based on its ability to maintain a monotonic and interpretable relationship between conflict severity and high-risk class probability. Transformations that produce erratic, flat, or counterintuitive trends are discarded. For example, using raw TTC values without transformation may lead to probability curves that plateau, failing to capture the behavioural urgency of imminent conflict. Conversely, inverse scaling (1/TTC) or other carefully selected forms can sharpen the response gradient and enhance behavioural realism. In this study, the reciprocal transformation 1/TTC appeared to provide a suitable balance between numerical stability and behavioural interpretability. It magnified the effect of low TTC values, where conflict is most critical, while preserving a smooth monotonic increase in the likelihood of high-risk classification. This choice was consistent with theoretical expectations that shorter TTC values elevate perceived threat and trigger risk-avoidant behaviour. More advanced forms of scaling or normalisation, such as exponentiation or adjustment relative to reaction time, may further enhance behavioural realism but are left for future work.

### 2.5.2. Primary Candidate Behavioural Thresholds (CBTs)

Since we hypothesise that CM probabilities follow a logistic shape, several points may be identified as candidate behavioural thresholds (CBTs). These points should not be viewed as definitive boundaries but rather as indicative regions where behavioural responses might shift. They are objectively identified from properties of the probability curve, such as its slope, inflection, or tail behaviour. Their interpretation must remain tentative given the absence of a foundational theory that prescribes exactly which curve features correspond to meaningful thresholds. Table 1 presents an illustrative set of these CBTs, along with their tentative interpretations and derivation rules.

In our framework, CBTs are mathematically identifiable points on the probability curve of CM that may correspond to behaviourally relevant transitions in driver responses to conflict. The suggestion of multiple thresholds along a continuum, rather than a single binary transition, appears consistent with evidence from several domains. This approach challenges simpler binary models and may better reflect how biological systems respond to graduated threats. A single threshold model might not capture the full range of dynamics observed in driving. Driver responses appear to span a continuum, from subtle awareness, through preparatory adjustments, to emergency action. Our CBTs, comfort zone, neutral point, crossover point, alertness point, and inflection point, o should therefore be seen as a possible attempt to identify and quantify such transitions within a unified mathematical framework.

Evidence from different disciplines appears to lend support to this multi-threshold perspective. In neuroscience, early studies suggested that the brain can shift between distinct processing modes (Buchsbaum & Silverman, 1968; J. Silverman et al., 1969). Later studies showed that the anterior cingulate cortex, a part of the brain that monitors conflict, adjusts control gradually in steps rather than through a simple on/off switch (Botvinick et al., 2001; Shenhav et al., 2013) . Cognitive psychology shows that behaviour is organised in layers. When tasks are easy, drivers rely on automatic routines, but as demands increase, higher-level supervisory control steps in to manage attention and decision-making (Braver, 2012; Norman & Shallice, 1986). Ethology shows that defensive behaviour unfolds in stages, from low alertness to full escape responses (Fanselow, 1994). Driving research also reveals a sequence: free driving, gas release, preparation, and braking, with comfort and panic zones marking boundaries (Horst & Hogema, 1993; Summala, 2007). Across these fields, behaviour has often been interpreted as progressing through stages rather than shifting at a single point. Building on this perspective, we outline below the candidate behavioural thresholds, each linked to a possible theoretical basis.

The *inflection point* is defined as the value of the conflict indicator where the derivative of the high-risk class probability reaches its maximum (the point of maximum slope). It represents the stage where behavioural sensitivity is highest. At this point, small changes in conflict severity indicate rapid shifts in response, marking the transition from preparatory behaviour to full evasive action. In neuroscience and psychophysics, this concept is captured through psychometric functions. These functions are widely used and known as S-shaped curves that link stimulus intensity to response probability. The slope of these curves reflects how abruptly behaviour changes. Low slopes indicate gradual adjustment, whereas steep slopes indicate sharp transitions. (Strasburger, 2001) showed that the maximum slope at the inflection point provides a robust, model-independent measure of perceptual change. (Gilchrist et al., 2005) emphasized slope as a key index of perceptual performance. In our framework, the inflection point marks where changes in CM probability occur most rapidly, meaning that small differences in the conflict indicator can strongly influence driver behaviour. To operationalize this, each conflict indicator is normalized by dividing by its mean, producing a dimensionless measure. The probability curve is then differentiated with respect to this normalized indicator, and the value at which the derivative reaches its maximum is taken as the inflection point:

$$CBT_{inflection} = \arg max\{(f'(CI))\}$$

Another candidate is the *crossover point*, which represents the region of greatest heterogeneity: some drivers begin defensive preparation, while others remain neutral. This point can be approximated as the value of the conflict indicator where the probability is 50%. In signal detection terms, it corresponds to the neutral criterion, where detection and non-detection are equally likely (Green & Swets, 1966). Behaviourally, it aligns with the preparation phase, marked by precautionary actions such as covering the brake, tightening grip, and increased fixation on the hazard (Horst & Hogema, 1993). In practice, it can be obtained by interpolating between the two probability values closest to 50%:

$$CBT_{crossover} = f_{class}(0.5)$$

The *alertness point* represents the upper bound of predominantly defensive behaviour. It can be defined as the value of the conflict indicator corresponding to the 95[th] percentile once the probability curve exceeds 50%. In practical terms, this is the point where nearly all drivers have initiated defensive action. The choice of the 95[th] percentile follows common practice in human factors. In this domain, design standards aim to accommodate most (but not all) individuals (Pheasant & Haslegrave, 2018). Based on this definition, the alertness point would generally be expected to fall below the crossover point:

$$CBT_{alertness} = Q_{0.95}(CI|f_{class}(CI) > 0.5)$$

The *neutral point* represents the stage where the probability of high-risk behaviour changes only minimally and approaches a baseline level. It marks the onset of awareness of potential conflict, but defensive responses are not yet triggered. The 0.05 slope threshold is consistent with psychophysical just-noticeable differences. At this stage, drivers shift from subconscious monitoring to conscious awareness, with increased mirror use and more focused visual attention (Underwood et al., 2003). Mathematically, it can be derived from the normalized conflict indicator and is defined as the value of the indicator at which the derivative of the probability equals 0.05:

$$CBT_{neutral} = \arg\{CI|f'_{class}(CI) = 0.05\}$$

Finally, the *comfort zone* corresponds to the asymptotic tail of the curve, where defensive behaviour is negligible and driving conditions reflect routine performance. In this state, drivers perceive little or no risk, safety margins are wide, and behaviour is mainly habitual. (Fuller, 2005) Task–Capability Interface model situates this region where task demand is far below capability, leading to automated, low-effort performance with broad safety margins. (Rasmussen, 1983) skill-based mode may also apply here, as driving is governed by automated routines that require minimal conscious control. These candidate thresholds may serve as plausible behavioural markers, though their robustness requires further evaluation. The next section outlines diagnostic criteria for assessing these thresholds and presents alternative methods that can be applied when primary estimates appear unreliable.

*Table 1 Illustrative set of potential CBTs – primary thresholds estimates*

| Behavioural Threshold | Interpretation | How It's Derived | Behavioural Basis/Theory |
|---|---|---|---|
| Inflection point | Point of max. behavioural sensitivity. Drivers may shift most rapidly toward defensive responses, showing heightened caution such as braking. | Inflection Point of logistic curve: maximum slope (steepest change in $P_{HR}$). | Psychophysics: slope of psychometric functions as an index of perceptual change (Gilchrist et al., 2005; Strasburger, 2001). |
| Crossover point | Boundary where defensive and neutral behaviours appear equally likely, suggesting a transition zone of mixed responses. | Defined where the probabilities of high-risk and low-risk membership are nearly equal ($P_{HR} = P_{LR} = 0.5$). | Signal detection theory: neutral criterion (Green & Swets, 1966). Behavioural preparation phase (Horst & Hogema, 1993). |
| Alertness point | Upper boundary of the high-risk zone ($P_{HR} > 0.5$). Drivers may still remain cautious, though defensive reactions seem to taper off. | 95% quantile of TTCs from observations with $P_{HR} > 0.5$. | Human factors practice: design for the 95th percentile to represent most individuals (Pheasant & Haslegrave, 2018). |
| Neutral point | Behaviour appears mostly neutral. Defensive responses become rare, and perceived risk seems minimal. | Cut-off where the probability of high-risk behaviour flattens. | Psychophysics: just noticeable differences as minimal perceptible change; attentional shift from subconscious to conscious (Underwood et al., 2003). |
| Comfort zone | Situation may not affect high-risk behaviour. Drivers tend to act habitually and may disregard possible risk cues. | Tail region of the logistic curve: asymptotic flattening of $P_{HR}$ | Task–Capability Interface model: demand far below capacity (Fuller, 2005). Rasmussen's (1983) skill-based behaviour mode. |

### 2.5.3. Alternate Candidate Behavioural Thresholds (CBTs)

Although the primary candidate thresholds outlined above may offer useful indicators of behavioural transitions, their reliability is not guaranteed. Certain curve characteristics can lead to unstable or implausible estimates. The probability curves derived from conflict indicators can sometimes produce unrealistic outcomes. This might happen when they are excessively steep, due to limited data, or deviate from expected mathematical properties. These issues could compromise the identification of meaningful thresholds. To address these challenges, we outline a diagnostic framework that considers potential sources of unreliability for each threshold type and points to alternative estimation methods that may be more suitable when standard approaches appear to fail.

Different threshold types seem to carry different vulnerabilities. Probability-based thresholds, such as inflection and crossover points, tend to be more sensitive to curve steepness. Distribution-based thresholds, such as the alertness point, neutral point, and comfort zone, appear more affected by sparse data in the distribution tails. Table 2 provides a summary of this framework, presenting diagnostic failure criteria for each threshold type alongside possible alternative methods. For instance, if an inflection point appears unstable, it may be approximated by locating the maximum curvature point through either a second derivative test or the Kneedle algorithm endpoints (Satopaa et al., 2011). Similarly, when crossover points become unreliable due to extreme steepness, segmented regression may provide a more stable alternative. For thresholds defined in distribution tails, such as the alertness point or comfort zone, adjustments based on percentiles or kernel density estimation could offer more robust estimates.

The logistic curve may be expressed as $P = \frac{1}{1+e^{-\mu(\frac{1}{x}-\frac{1}{x_o})}}$, where $x$ is the conflict indicator value (*e.g.*, TTC in seconds), $x_o$ is the conflict indicator value where the probability is 0.5 (the 50% threshold), and $\mu$ controls the steepness of the transition. We tentatively adopt $|\mu| > 2$ as a practical indicator of potential instability. This is because the middle part of the probability curve may represent the range where values rise from about 12% to 88%. As $\mu$ increases, this range gets squeezed into a narrower interval. To illustrate, when $\mu = 1$, the 12%-88% transition spans about ±4 units from $x_o$, reflecting a gradual change. When $\mu = 2$, the transition becomes about ±2 units, indicating a much sharper behavioural shift. When $\mu = 4$, the transition is restricted to just ±1 unit, making the change extremely abrupt.

Thus, higher values of $\mu$ correspond to increasingly steep curves where the probability of high-risk behaviour changes almost instantaneously with small shifts in the conflict indicator. For instance, if we consider a TTC-based indicator, the behavioural transition is compressed into only two seconds. Since this duration is comparable to or faster than typical human response times (0.75–1.5s; Green, 2000), these transitions may represent behavioural changes that are too abrupt to support stable threshold extraction. Hence, probability-based thresholds could become highly sensitive to data variations, with even small measurement errors shifting estimates considerably. In addition to curve-based diagnostics, physical reasonableness should also be considered. Prior research suggests that TTC thresholds could fall within about 0.5–6 seconds, consistent with observed human reaction times and traffic dynamics (Hayward, 1972; Hydén, 1987; Van der Horst, 1991; Vogel, 2003; Noy, 1997).

Apart from these primary diagnostics, several complementary methods could also be considered when standard thresholds are unreliable. For example, the Kneedle algorithm detects the point of maximum curvature where the probability curve deviates most from a straight line connecting its endpoints (Satopaa et al., 2011). This may help when inflection points are poorly defined. Another approach is the segmented regression, which fits two connected linear segments with the breakpoint as the behavioural threshold (Muggeo, 2003). This could also provide stability when steep slopes destabilise probability-based thresholds. For distribution-based thresholds, percentile adjustments may be appropriate. If the 95th percentile is supported by too few observations, the 90th percentile may yield a more robust estimate (Wilcox, 2011). In addition, the kernel density estimation (KDE) can highlight the most likely threshold by locating peaks in smoothed distributions of high-risk observations (B. W. Silverman, 2018). Finally, change-point detection methods such as the PELT algorithm can identify where the probability curve shifts from steep decline to relative stability. This may indicate the point at which defensive behaviour levels off (Killick et al., 2012).

In addition to these methods, it may also be useful to conduct a resampling stability check of identified CBTs. One possible approach is to take random subsamples of the dataset, re-estimate the model, and re-identify the thresholds. If the thresholds remain within a narrow range across subsamples, this could provide additional confidence. If they vary widely, this may signal instability and the need for further adjustment. The diagnostic criteria in Table 2 ($\mu > 2$ and $\mu > 3$) may reflect different levels of instability. When $\mu > 2$, the 12%-88% probability transition occurs within ±1 unit of $x_0$, which implies that the crossover point is highly sensitive to measurement errors. For instance, a 0.5-second variation in TTC could shift the estimated threshold by several seconds. When $\mu > 3$, this transition narrows further to around ±0.67 units, which can lead to numerical instability in the second derivative calculations used to identify inflection points.

*Table 2 Diagnostic framework for CBTs reliability – alternative thresholds estimates*

| Threshold Type | Rationale/Theoretical Basis | Curve Sensitivity | Diagnostic Failure Criteria | Alternative Method When Failed |
|---|---|---|---|---|
| Inflection point | Point of maximum behavioural sensitivity where drivers show most rapid response changes; aligns with critical reaction zones (Van der Horst, 1990; Hirst & Graham, 1997) | • High sensitivity to $\mu$<br>• Unstable when $\mu > 3$ | • $\mu > 3$ (unstable inflection)<br>• Poor second derivative definition<br>• Non-monotonic curve | Maximum curvature point using:<br>• Second derivative<br>• Or Kneedle algorithm |
| Crossover point | 50% probability threshold; could be viewed as the majority region where drivers are equally split between high vs. low risk | • Very high sensitivity to $\mu$<br>• Critical failure when $\mu > 2$ | • Value > 15s (extrapolation issue)<br>• $\mu > 2$ (too steep) | Segmented regression breakpoint (Muggeo, 2003) |
| Alertness point | • Statistical upper bound (95% quantile) of high-risk observations<br>• Represents tail of defensive behaviour distribution | • Low sensitivity to $\mu$<br>• Very high sensitivity to data sparsity | • Value < inflection point (illogical ordering)<br>• n < 30 in tail region<br>• Non-monotonic with other thresholds | 95% quantile or kernel density peak |
| Neutral point | • Transition where probability curve flattens<br>• Indicates defensive responses become rare and risk perception minimal | • Medium sensitivity to $\mu$<br>• High sensitivity to data coverage | • No flattening observed<br>• Derivative < 0.05 throughout<br>• Beyond data range | First point where derivative < 0.05 or change-point detection |
| Comfort zone | • Asymptotic tail region where defensive behaviour becomes negligible<br>• Represents habitual driving conditions | • Low sensitivity to $\mu$<br>• Very high sensitivity to tail sparsity | • Value > 30s (sparsity artifact)<br>• n < 20 beyond this point | 99% quantile of full distribution |

### 2.5.4. Validation of Candidate Behavioural Thresholds (CBTs)

To determine whether a CBT of a conflict indicator corresponds to a meaningful behavioural shift, two validation approaches are proposed. The first relies on the Chow test. The second involves estimating alternative LC-LK model specifications that differ in how the conflict indicator enters the CM function, once as a continuous term, once as a binary threshold variable, and once as a combination of both. The Chow test provides a statistically robust method for evaluating whether decision-making parameters differ significantly on either side of a proposed threshold (Toyoda, 1974). This test is particularly useful when behavioural changes are hypothesised to emerge from variations in perceived conflict intensity. To implement the test, a candidate threshold (*i.e.*, crossover point) is defined, and the dataset is divided into two subsets using this threshold point value. Group 1 includes observations where the indicator is less than or equal to the threshold. Group 2 includes observations where the indicator exceeds the threshold. Separate models with identical utility specifications are estimated for each group, Model I on Group 1 and Model II on Group 2. The Chow statistic is then calculated as:

$$LR_{Chow} = -2 \times (LL_{full} - (LL_1 + LL_2))$$

where $LL_{full}$ is the log-likelihood (LL) of the model that includes the undivided dataset, $LL_1$ and $LL_2$ are the LL of Models I and II, respectively. This test statistic follows a chi-squared distribution with *df* equal to the number of estimated parameters. A statistically significant result indicates a structural change in model parameters, confirming that the behavioural response differs across the threshold and thus validating its relevance. The second approach evaluates the threshold by estimating multiple versions of the LC-LK with alternative specifications of incorporating the conflict indicator in the CM:

- In the baseline model, the indicator enters the CM function as a continuous transformation (*e.g.*, 1/TTC).
- In the threshold model, a binary variable (*e.g.*, TTC ≤ threshold) is used instead of the continuous term.
- A third specification combines both: the binary threshold indicator and its interaction with the continuous term.

Model comparisons are conducted using information criteria (*e.g.*, AIC, BIC) and Likelihood Ratio Tests (LRTs) to determine whether threshold-based terms enhance model performance. If the model that includes the binary indicator and its interaction with the continuous variable provides a statistically significant improvement over the baseline, this suggests that the behavioural influence of the conflict indicator is not constant across its range. This, in turn, supports the interpretation of the proposed threshold as a potential tipping point in behavioural response. This method offers an alternative to the Chow test when data imbalance or sample size limitations constrain the feasibility of re-estimating models on partitioned subsets. In some cases, as observed in this study, selecting a candidate threshold may result in a small number of observations either above or below that threshold. This imbalance can limit the feasibility of re-estimating models on partitioned subsets. To address this, an alternative is to estimate versions of the LC-LK in which only an ASC is specified in the CM probability. An LRT can then be conducted against the baseline model to examine whether including the binary threshold term (*e.g.*, indicator ≤ threshold) provides a statistically meaningful improvement in fit. Evidence of such improvement would suggest that the threshold captures a behavioural distinction, even when sample size constraints make the Chow test less reliable.

## 3. Results and Discussions

### 3.1. Model Description

We apply the proposed framework to two conflict indicators: TTC and modified time-to-collision (MTTC2). Both indicators are only defined when two vehicles are on a potential collision course. TTC assumes constant velocity when projecting future trajectories. The original modified time-to-collision (MTTC), introduced by Ozbay et al., (2008), is defined as the time remaining until two vehicles collide if they continue on their current paths, under the assumption of constant acceleration in the direction of velocity. In this paper, we extend that formulation by considering the full vector definition of acceleration, without restricting it to the velocity direction. This extension is referred to as MTTC2. Since vehicles at roundabouts typically travel in curvilinear trajectories, MTTC2 could provide a more realistic representation of conflict dynamics compared to both TTC and the original MTTC. TTC is included as the baseline (simplest case), while MTTC2 captures richer dynamics by incorporating vector-based acceleration. Formally, MTTC2 is obtained by:

$$MTTC2 = min\left\{t > 0 \middle| \left\|r_o + \Delta vt + \frac{1}{2}at^2\right\| = 0\right\}$$

where $r_o = x_2(0) - x_1(0)$ is the initial relative position, $\Delta v = v_2 - v_1$ is the relative velocity, and $\Delta a = a_2 - a_1$ is the relative acceleration, capturing both the magnitude and direction of the difference in accelerations between the two vehicles. Figure 2 presents an illustration of two vehicles movement extrapolation using non-constant acceleration, and the resulting collision point using this definition. Following Laureshyn et al (2010), for each vehicle pair we check all 32 possible corner to side and side to corner combinations (assuming rectangular shapes). The minimum positive TTC and MTTC2 among these combinations is taken as the effective collision time.

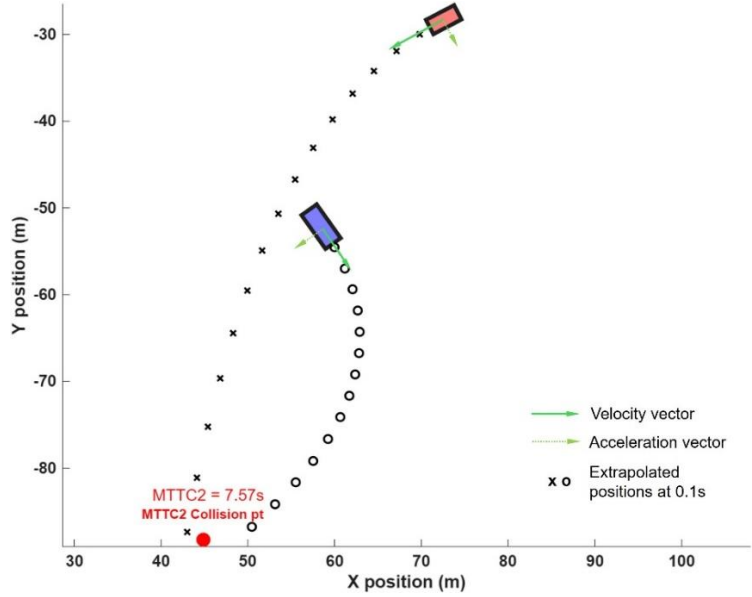

*Figure 2 Modified time-to-collision (MTTC2) illustration for two vehicles considering vector-based acceleration.*

The open roundabout drone (OpenDD) dataset is utilized in this paper (Breuer et al., 2020). From this dataset, roundabout site number 2, a single-lane roundabout, is selected for applying the proposed framework. A total of 46,120 observed choices made by passenger cars are extracted at 1-second intervals. At each time step, every car is interacting with a number of surrounding vehicles, and for each car the conflict indicators with the lowest values are identified: TTC and MTTC2. From this, a subset of 2,045 interaction observations is obtained for TTC and 1,935 for MTTC2, each representing the minimum value of the corresponding indicator for a passenger car at a given time step.

The systematic utility specification for the CM specification follows a similar structure introduced in Section 2.3.1. The systematic utility $V_{nit}^s$ specification in the CC probability for each class, where $HR$ represents the high-risk class, and $LR$ is the low-risk class is defined by:

$$V_{nit}^s = \begin{cases} \gamma_D I_{Dnit} CAI_{nkt}^{Front} + \gamma_T \left[ I_{nit}^{Right} CAI_{nkt}^{Left} + I_{nit}^{Left} CAI_{nkt}^{Right} \right] + \gamma_{dp} dp_{nit} & s = HR \\ +\eta_{Speed}^{123} + e^{\alpha_1^{Spd}} \eta_{Speed}^{456} + e^{\alpha_2^{Spd}} \eta_{Speed}^{789} + \eta_{Turn}^{147} + e^{\alpha_1^{Trn}} \eta_{Turn}^{258} + e^{\alpha_2^{Trn}} \eta_{Turn}^{369} & \\ \\ \delta_D I_{Dnit} + \delta_A I_{Anit} + \delta_T I_{Tnit} + \delta_C I_{Cnit} + \delta_{dp} dp_{nit} & s = LR \\ +\eta_{Speed}^{123} + e^{\alpha_1^{Spd}} \eta_{Speed}^{456} + e^{\alpha_2^{Spd}} \eta_{Speed}^{789} + \eta_{Turn}^{147} + e^{\alpha_1^{Trn}} \eta_{Turn}^{258} + e^{\alpha_2^{Trn}} \eta_{Turn}^{369} & \end{cases}$$

Here, the $CAI_{nkt}^{Front}$ represents the directional collision angle intensities associated with the lowest value of the conflict indicator used in CM, more details can be found in (Al-Haideri et al., 2025a). These directional intensities capture the effect of turning maneuvers and deceleration behaviour when there is a vehicle on the left or on the right side or in front of the driver serving as a collision avoidance mechanism to capture an evasive action in the high-risk class. The $dp_{nit}$ captures the deviation of driver from an ideal trajectory or a dynamic attraction to destination measured from the centroid of each choice, more details can be found in (Al-Haideri et al., 2025a). The dummy $I_{Dnit}$ is equal to 1 if the driver selects deceleration alternatives (1, 2, 3 in Figure 1). The dummy $I_{Anit}$ is equal to 1 if the driver selects acceleration alternatives (7, 8, 9 in Figure 1). The dummy $I_{nit}^{Right}$ is equal to 1 if the driver selects right-turn alternatives (3, 6, 9 in Figure 1). The dummy $I_{nit}^{Left}$ is equal to 1 if the driver selects right-turn alternatives (1, 4, 7 in Figure 1). The dummy $I_{Tnit}$ is equal to 1 if the driver selects turning alternatives (3, 6, 9, 1, 4, 7 in Figure 1). The dummy $I_{Cnit}$ is equal to 1 if the alternative is not occupied by another vehicle or falls on the roundabout's physical geometry.

The error structure specification represents a critical modelling decision balancing behavioural realism with statistical identification. We tested various error structures ranging from fully flexible to highly restricted specifications. The final model employs two base error components with four estimated scale parameters. This structure achieved robust convergence across all models while more complex specifications consistently failed. This convergence success provides empirical support for the chosen structure, reflecting actual patterns in how drivers navigate roundabouts. The behavioural interpretation of what these components represent becomes clearer through the estimation results of the parameters, which we discuss in detail in the next sections. The two factor structure captures fundamental dimensions of roundabout navigation behaviour as revealed by the data. The speed error component ($\eta_{Speed}$) could represent an underlying driver state, such as urgency, aggressiveness, or confidence, that simultaneously affects all speed related alternatives. When this latent factor takes positive values, it increases the utility of alternatives within the same speed ring, though with different intensities determined by its estimated scales. The directional error component ($\eta_{Turn}$) also captures systematic preferences for turning maneuvers. This creates correlation among alternatives sharing directional characteristics. This decomposition reflects how drivers at roundabouts process alternatives as combinations of two fundamental decisions (speed adjustment and directional choice), rather than evaluating nine independent alternatives.

The scale parameters offer tentative behavioural insights. The deceleration and left turn scales are normalized to one for identification purposes. The keep speed scale ($e^{\alpha_1^{Spd}}$) and acceleration scale ($e^{\alpha_2^{Spd}}$) might indicate how sensitive these alternatives are to the underlying speed preference. Though whether this represents true behavioural heterogeneity or merely captures unobserved factors, this remains uncertain. Similarly, the straight ($e^{\alpha_1^{Trn}}$) and right turn ($e^{\alpha_2^{Trn}}$) scales might suggest varying importance of directional preferences. Larger scales could indicate greater heterogeneity in driver preferences. Maybe alternatives becoming more selective and chosen only by drivers with strong preferences, though this interpretation should be considered provisional. The parsimony of this specification appeared essential for achieving identification. However, we cannot definitively say that a more complex structure might succeed with different data or estimation approaches. With only two base errors and four scale parameters, the model captures correlation patterns while maintaining numerical stability. More complex error structures failed to converge, possibly due to over-parameterization creating flat likelihood, weak identification from insufficient data variation, or multicollinearity among error components. However, it could be also possible that with larger samples or different optimization algorithms, richer structures might be identifiable.

### 3.2. Estimation Results

Model estimation was performed using maximum simulated likelihood estimation coupled with an expectation-maximization algorithm in GAUSS 25 (Aptech Systems, 2025). In this section, we implement model estimation and diagnostics to examine differences between using two data subsets. Table 3 reports the LC-LK estimation results for TTC, comparing the full dataset with the interaction subset. Table 4 presents the corresponding results for MTTC2, also estimated on both the full dataset and the interaction subset. All estimated parameters in the CM and CC are statistically significant at the 5% confidence level and have the expected signs, except for the deceleration in the low-risk class and scale for error of turn right cone. All coefficient signs remain consistent across both tables and data subsets, with the exception of the distance to path parameter in the high-risk class, which will be discussed bellow. The estimation results from Tables 3 and 4 reveal fundamental differences in driver behaviour between general roundabout navigation and active conflict situations, providing crucial insights into how drivers perceive and respond to collision risks at single-lane roundabouts.

### 3.2.1. Class Membership (CM) Estimates Interpretation

The baseline propensity for high-risk class membership ($\beta_{ASC}$) exhibits completely different effects across datasets. In Model A (TTC interaction subset), the strongly negative ASC ($\beta_{ASC}$= -1.87) indicates that drivers are default to routine (low-risk) behaviour when consciously engaged with other vehicles (TTC is high). Model C (MTTC2 interaction subset) shows even stronger baseline preference for low-risk behaviour ($\beta_{ASC}$= -2.24). However, Model B (TTC full data) shows the baseline parameter approaching zero ($\beta_{ASC}$= -0.291). While in Model D (MTTC2 full data), it is nearly zero ($\beta_{ASC}$= -0.051). This near-zero baseline in the full data models implies approximately 40-45% probability of exhibiting high-risk class characteristics during routine driving. At a single-lane roundabout, this may suggest that drivers maintain constant vigilance even without immediate conflicts. Probably potentially scanning for entering vehicles, monitoring curved sight lines, or preparing for unexpected pedestrian crossings. Such behaviours might be essential for roundabout safety.

The conflict indicator coefficients ($\beta_{CI}$) in both Tables 3 and 4 demonstrate successful capture of behavioural transitions. Model A yields $\beta_{CI}$= 3.707 for TTC interactions. Model B corresponds to $\beta_{CI}$= 1.830 for the full dataset, a reduction of approximately 50% than Model A. Similarly, Model C (MTTC2 interactions) has $\beta_{CI}$= 3.279 compared to Model D's $\beta_{CI}$= 1.330. This amplification in interaction only models could make behavioural sense. Specifically, when drivers are actively engaged in conflicts at roundabout entry or within the circulatory roadway, they become hypersensitive to collision indicators, with the conflict measure dominating their behavioural state. The slightly higher coefficients for TTC compared to MTTC2 across all models suggests drivers at roundabouts may better perceive constant-velocity projections, possibly because the relatively stable speeds in single-lane roundabouts make acceleration based projections less intuitive.

The error structure parameters reveal compelling patterns about behavioural heterogeneity. In Model B (TTC full data), the acceleration scale parameter ($e^{3.075} \approx 21.5$) indicates extreme variance in acceleration choices, only highly aggressive or confident drivers accelerate within roundabouts during normal conditions. This could make perfect sense behaviourally, as acceleration in a roundabout is generally discouraged and potentially dangerous. However, in Model A (TTC interactions), this scale drops dramatically ($e^{1.007} \approx$ 2.7), suggesting acceleration becomes a more standardized evasive maneuver during conflicts, perhaps to escape a developing collision scenario by accelerating through a gap. The same pattern appears in Models C and D for MTTC2. Notably, the right-turn scale parameters are insignificant in both interaction models (p=0.738 in Model A; p=0.703 in Model C), suggesting that during active conflicts, right-turn heterogeneity disappears as drivers' alternatives become constrained by immediate threats.

### 3.2.2. Class-Specific Estimates Interpretation

The class-specific utility parameters in Tables 3 and 4 reveal distinct behavioural traits. In the low-risk class, Model B shows strong negative coefficients for deceleration ($\delta_D$ = -10.123), confirming that routine drivers actively avoid unnecessary slowing to maintain efficient roundabout flow. This behaviour is essential for roundabout functionality, where unnecessary braking disrupts the continuous flow design. The positive acceleration coefficients in interaction Models A ($\delta_A$= 1.318) and C ($\delta_A$ = 1.644, p<0.001) versus negative coefficients in full data Models B and D suggest that routine acceleration serves different purposes: gap-taking behaviour during interactions versus generally discouraged behaviour in normal flow. The high-risk class parameters demonstrate strong defensive behaviours appropriate for roundabout conflicts. Deceleration with frontal collision intensity shows large positive coefficients in all models, but notably doubles in interaction subsets (Models A: β = 26.704; Model C: β = 30.675) compared to full datasets (Model B: $\gamma_D$ = 12.814; Model D: $\gamma_D$ = 12.785). This amplification makes behavioural sense, when facing an immediate frontal threat at roundabout entry or from a vehicle cutting across the circulatory roadway, emergency braking becomes the primary defensive response.

The path deviation parameters reveal interesting behavioural adaptations. In low-risk conditions, all models show negative coefficients for distance from path, indicating routine preference for staying on course. However, the high-risk class in interaction models (A and C) shows positive path deviation coefficients ($\gamma_{dp}$=0.746 and $\gamma_{dp}$=0.476 respectively), suggesting willingness to deviate from the ideal trajectory when avoiding collisions, a critical safety behaviour in roundabouts where the curved geometry provides lateral escape options. The reversal to negative coefficients in full data models (B and D) suggests that general defensive driving still prioritizes path adherence when immediate threats are absent. The turning behaviour parameters provide insights into directional responses. In the low-risk class, the extremely high positive coefficients for turning ($\gamma_T$=11.514 in Model A; $\gamma_T$=12.555 in Model C) during interactions suggest that routine turning becomes strongly preferred when navigating around other vehicles, consistent with the yielding and merging behaviours essential to roundabout operations. The collision-dependent turning in the high-risk class intensifies dramatically in interaction models (4.181 in Model A; 5.831 in Model C) compared to full data models, indicating that evasive steering becomes a primary defensive strategy when facing lateral collision threats from merging or exiting vehicles.

*Table 3 Model estimation results using the full dataset and interaction subset for TTC.*

| Model | | Model A (Interaction Subset) | | Model B (Full Data) | |
|---|---|---|---|---|---|
| Parameter | | Estimates (SE) | p-value | Estimates (SE) | p-value |
| *Class Membership (CM) Probability (High-Risk Class)* | | | | | |
| $\beta_{CI}$ (1/TTC) | | 3.707 (0.316) | 0.000 | 1.830 (0.122) | 0.000 |
| $\beta_{ASC}$ | | -1.87 (0.156) | 0.000 | -0.291 (0.016) | 0.000 |
| *Conditional Choice (CC) Probability* | | | | | |
| *Class-Specific Coefficients* | | | | | |
| $\delta_D$ (dec.) | Low-Risk Class | -0.561 (0.493) | 0.255 | -10.123 (0.556) | 0.000 |
| $\delta_A$ (acc.) | | 1.318 (0.332) | 0.000 | -2.890 (0.231) | 0.000 |
| $\delta_C$ (alternative is available) | | -2.354 (0.340) | 0.000 | -2.188 (0.082) | 0.000 |
| $\delta_p$ (distance to path from choice cell) | | -0.573 (0.060) | 0.000 | -1.246 (0.022) | 0.000 |
| $\delta_T$ (turning left or right) | | 11.514 (2.709) | 0.000 | 1.969 (0.047) | 0.000 |
| $\gamma_D$ (dec. with frontal collision intensity) | High-Risk Class | 26.704 (4.496) | 0.000 | 12.814 (1.189) | 0.000 |
| $\gamma_{dp}$ (distance to path from choice cell) | | 0.746 (0.174) | 0.000 | -0.507 (0.015) | 0.000 |
| $\gamma_T$ (turn left/right with collision intensities) | | 4.181 (0.512) | 0.000 | 0.639 (0.109) | 0.000 |
| *Generic Coefficients* | | | | | |
| $e^{\alpha_1^{Spd}}$ (Scale for error of keep speed ring (4-5-6)) | | 1.884 (0.170) | 0.000 | 2.069 (0.049) | 0.000 |
| $e^{\alpha_2^{Spd}}$ (Scale for error of accelerate speed ring (7-8-9)) | | 1.007 (0.247) | 0.000 | 3.075 (0.043) | 0.000 |
| $e^{\alpha_1^{Trn}}$ (Scale for error of keep direction cone (2-5-8)) | | 3.231 (0.202) | 0.000 | 0.973 (0.025) | 0.000 |
| $e^{\alpha_2^{Trn}}$ (Scale for error of turn right cone (3-6-9)) | | -0.042 (0.125) | 0.738 | 0.613 (0.025) | 0.000 |
| Mean LL | | -1.925 | | -1.956 | |
| Sample size | | 2,045 | | 46,120 | |

*Table 4 Model estimation results using the full dataset and interaction subset for MTTC2.*

| Model | | Model C<br>(Interaction Subset) | | Model D<br>(Full Data) | |
|---|---|---|---|---|---|
| Parameter | | Estimates (SE) | p-value | Estimates (SE) | p-value |
| **Class Membership (CM) Probability (High-Risk Class)** | | | | | |
| $\beta_{CI}$ (1/MTTC2) | | 3.279 (0.295) | 0.000 | 1.330 (0.096) | 0.000 |
| $\beta_{ASC}$ | | -2.244 (0.198) | 0.000 | -0.051 (0.015) | 0.000 |
| **Conditional Choice (CC) Probability** | | | | | |
| *Class-Specific Coefficients* | | | | | |
| $\delta_D$ (dec.) | Low-Risk Class | -1.059 (0.670) | 0.114 | -10.101 (0.592) | 0.000 |
| $\delta_A$ (acc.) | | 1.644 (0.380) | 0.000 | -2.810 (0.254) | 0.000 |
| $\delta_C$ (alternative is available) | | -2.734 (0.433) | 0.000 | -2.187 (0.086) | 0.000 |
| $\delta_p$ (distance to path from choice cell) | | -0.616 (0.065) | 0.000 | -1.320 (0.025) | 0.000 |
| $\delta_T$ (turning left or right) | | 12.555 (3.173) | 0.000 | 1.989 (0.043) | 0.000 |
| $\gamma_D$ (dec. with frontal collision intensity) | High-Risk Class | 30.675 (5.324) | 0.000 | 12.785 (1.056) | 0.000 |
| $\gamma_{dp}$ (distance to path from choice cell) | | 0.476 (0.162) | 0.004 | -0.515 (0.013) | 0.000 |
| $\gamma_T$ (turn left/right with collision intensities) | | 5.831 (0.866) | 0.000 | 0.630 (0.101) | 0.000 |
| *Generic Coefficients* | | | | | |
| $e^{\alpha_1^{Spd}}$ (scale for error of keep speed ring (4-5-6)) | | 2.129 (0.166) | 0.000 | 1.857 (0.048) | 0.000 |
| $e^{\alpha_2^{Spd}}$ (scale for error of acc. speed ring (7-8-9)) | | 1.449 (0.212) | 0.000 | 2.978 (0.046) | 0.000 |
| $e^{\alpha_1^{Trn}}$ (scale for error of keep direction cone (2-5-8)) | | 3.379 (0.229) | 0.000 | 0.879 (0.025) | 0.000 |
| $\alpha e^{\alpha_2^{Trn}}$ (scale for error of turn right cone (3-6-9)) | | 0.045 (0.117) | 0.703 | 0.616 (0.027) | 0.000 |
| Mean LL | | 1.938 | | -1.957 | |
| Sample size | | 1,935 | | 46,120 | |

### 3.3. Primary Candidate Behavioural Thresholds (CBTs) for TTC and MTTC2

We illustrate the CBTs for the TTC and MTTC2 indicators using the full dataset, as shown in Table 5. The neutral point and comfort zone appear at long TTCs (9.5s), which seems to result from the logistic tail. Behaviourally, the crossover and alertness points may be more meaningful, as they appear to capture where most drivers tend to shift their behaviour. The neutral point is perhaps better interpreted as a theoretical boundary where defensive behaviour becomes negligible, rather than a realistic operational threshold. It is also noteworthy that MTTC2 consistently yields lower threshold values than TTC. This could be related to the extrapolation method in each indicator. The TTC assumes constant velocity, which may make potential conflicts appear to occur further in the future. The MTTC2 employs non-constant acceleration, which may reduce the projected time horizon and make conflicts appear more imminent. The EVT/GPD statistical thresholds fall within the range bounded by the inflection and alertness points. This suggests some alignment between statistical rarity and behavioural sensitivity. At the same time, the two perspectives are not equivalent and should be interpreted as complementary rather than interchangeable. Behaviourally, this tentative overlap may indicate that drivers' heightened caution and responsiveness tend to arise in the same region where extreme and rare conflicts are concentrated statistically.

*Table 5 Candidate behavioural thresholds (CBTs) for TTC and MTTC2*

| Data | Interaction subset | | Full data | |
|---|---|---|---|---|
| Conflict indicator threshold | TTC (s) | MTTC2 (s) | TTC (s) | MTTC2 (s) |
| **Behavioural thresholds** | | | | |
| Inflection point | 1.13 | 0.96 | 0.83 | 0.63 |
| Alertness point | 1.95 | 1.43 | 4.98 | 5.84 |
| Crossover point | 1.99 | 1.46 | 6.29 | 34.41 |
| Neutral point | 7.68 | 4.78 | 6.39 | 4.00 |
| Comfort zone | >7.68 | >4.78 | >6.39 | >4.00 |
| **Statistical threshold** | | | | |
| GPD threshold from EVT | 1.57 | 1.10 | 1.57 | 1.10 |

From Table 5, we select two thresholds, the crossover point and the inflection point, for further visualisation, and apply diagnostic checks to assess whether a behavioural shift could be occurring. Figure 3 illustrates the estimated CM probabilities as a function of each conflict indicator. The figure compares two models: one estimated using the interaction subset and the other using the full dataset. Across both indicators and datasets, the probability of being classified as high-risk appears to decline as the conflict indicator increases. This trend appears broadly consistent with the expectation that shorter temporal proximities are more likely to be associated with heightened risk perception. This in turn may lead to evasive actions. At longer TTC or MTTC2 values, the probability of defensive behaviour tends to flatten toward a baseline level of caution.

When comparing the two models, some differences in the steepness and curvature of the probability functions can be observed. In the interaction subset, the decline in CM probability appears sharper. Here, the crossover points occur around 2s for TTC and 1.5s for MTTC2. This may suggest that when focusing only on collision course events, the model puts more emphasis on the transition between high-risk and low-risk behaviour. In contrast, the full dataset models produce smoother declines. In these graphs, the crossover points are about 6s for TTC and 34s for MTTC2. The latter may be implausible given that it implies that 50% of drivers would still be in the high-risk class at very long conflict times, as will be discussed in the following section.

Figure 4 presents the relationship between the derivatives of the CM probabilities with respect to TTC and MTTC2, highlighting the respective inflection points. These plots point to regions where behavioural sensitivity to the conflict indicator may be most pronounced. Across both indicators and datasets, the strongest rate of change in high-risk probability appears to occur at relatively short values of the conflict indicators. The inflection points for TTC are estimated at 1.1s for the interaction subset and 0.8s for the full dataset. For MTTC2, the values are 1.0s and 0.6s, respectively. These results suggest that the steepest behavioural transitions are focused on situations of imminent conflict, when temporal proximity is very limited. The similarity of inflection points across indicators and dataset definitions indicates that the point of maximum behavioural sensitivity seems relatively stable, lending more credibility to this threshold compared to the crossover point.

Both dataset types provide useful insights but emphasise different facets of driver behaviour. The full dataset incorporates non-collision and free-flow situations and tends to show steeper transitions in the high-risk probability curve. In contrast, the interaction subset, focusing solely on collision-course events, produces smoother transitions between low- and high-risk states. These differences suggest that each approach offers a complementary perspective on how behavioural thresholds emerge. Rather than selecting one, we present both for comparison and suggest that future work, with further validation and sensitivity testing could clarify which dataset better captures real-world driver responses.

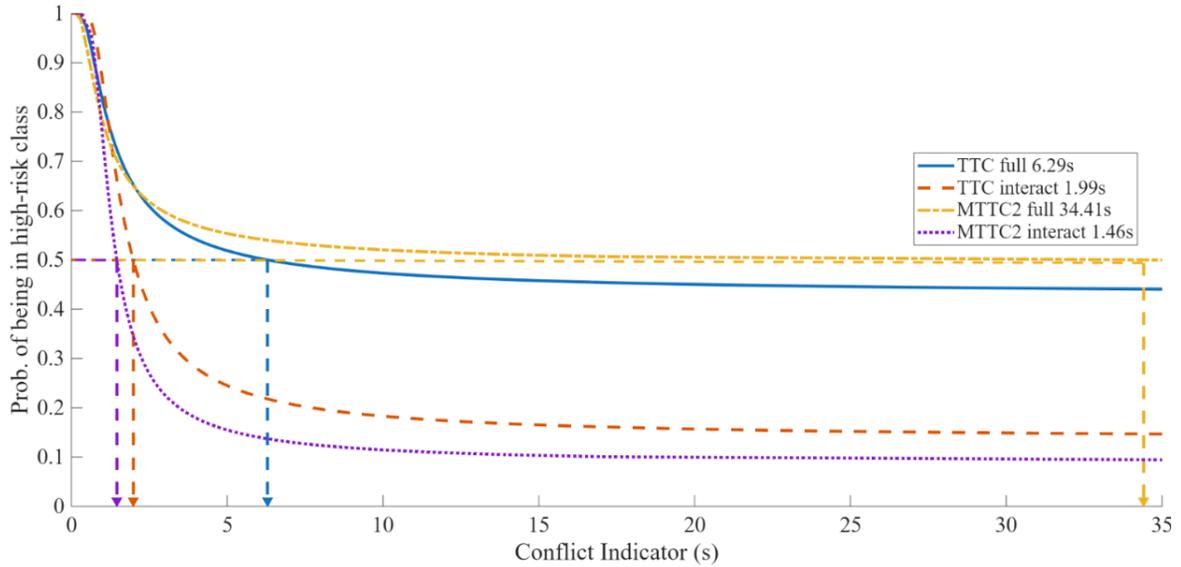

*Figure 3 Crossover points for the TTC and MTTC2 conflict indicators*

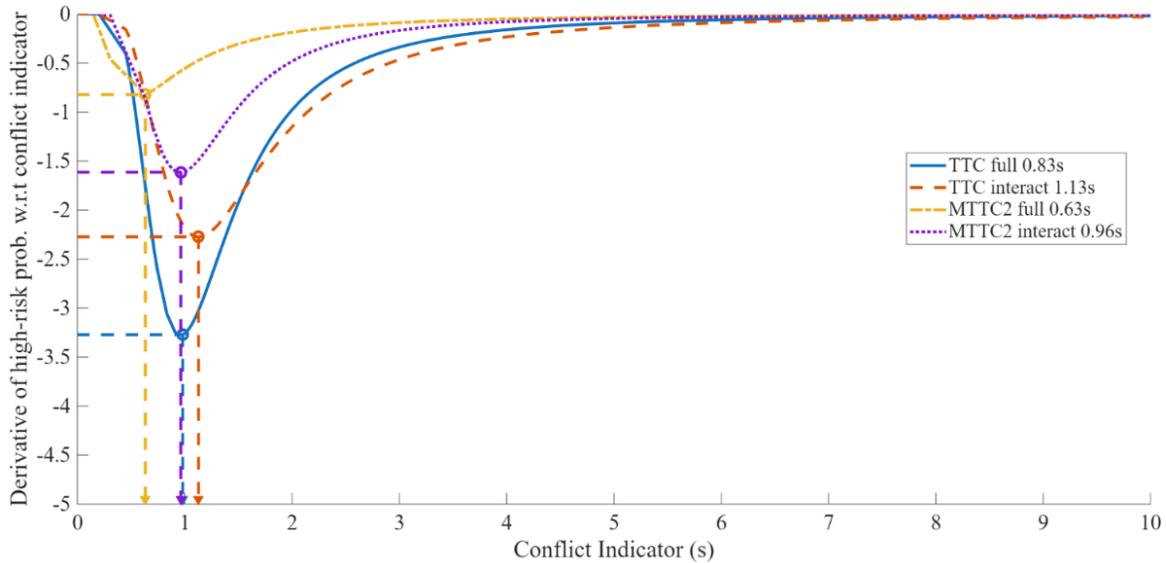

*Figure 4 Inflection points for the TTC and MTTC2 conflict indicators*

### 3.4. Alternate Candidate Behavioural Threshold (CBT) for MTTC2

As outlined in Section 2.5.3, the MTTC2 crossover identified from the full data raises diagnostic concerns. The probability curve for MTTC2 shows a long flat tail. The 50% crossover falls in this region, where the slope is very low. Under such conditions, the threshold may be unstable and sensitive to small variations. This potential instability suggests that the crossover should not be relied upon for MTTC2 in the full dataset. Alternative methods such as segmented regression may provide more robust estimates. In the interaction subset, the crossover was 1.46s, which seemed plausible. This difference may suggest that non-collision events altered the response pattern. To explore this, we applied segmented regression. The two-segment model gave a breakpoint at 2.50s (95% CI: 2.47–2.52). The three-segment model added more detail, with breakpoints at 1.4s and 4.62s, as shown in Figure 5.

These two breakpoints may align with other thresholds. The first at 1.40s was close to the EVT/GPD threshold of 1.1s for MTTC2. Both fall in a range linked with extreme events. This overlap may indicate that the shift at 1.40s coincides with a statistical boundary. The second breakpoint at 4.62s was near the 4s neutral point from our behavioural analysis. This similarity could mean that the model captured the same transition. The divergence between datasets may explain why MTTC2 needs correction in a roundabout context. In the interaction subset, both TTC and MTTC2 had interpretable thresholds (1.99s and 1.46s). The 1.46s crossover was also close to the segmented breakpoint (1.40s) and EVT threshold (1.10s). These overlaps may suggest that a core behavioural boundary was captured by different methods. In these cases, drivers seemed to respond to actual conflicts with clear transitions.

In the full dataset, many events were routine circulation. These included following at safe spacing, slowing to exit, or accelerating to merge. TTC often produced large or undefined values, which seemed consistent with safety. MTTC2, however, could have still given finite values because roundabout geometry involves continuous acceleration. These accelerations may have created apparent convergence even when vehicles were safe. For example, a slowing vehicle at exit and a following vehicle at constant speed could yield moderate MTTC2 values despite no conflict. This likely distorted the probability curve and produced the flat tail that caused the 34s crossover. Roundabouts may make this problem worse. Curved paths require frequent acceleration changes. MTTC2 interprets these as conflicts. Segmented regression may be useful for MTTC2 in such datasets. TTC, by contrast had a monotonic decline in risk and a stable threshold at 6.29s. MTTC2, with its acceleration term, introduced more uncertainty. Even at 30s, it may have signalled potential risk.

The three-segment model helped address this. It also hinted at behavioural phases that might matter in roundabouts. The first segment (0–1.4s) showed a steep drop. This may reflect emergency cases such as failure-to-yield. The second (1.40–4.62s) showed a moderate decline. This may reflect drivers distinguishing conflicts from normal navigation. The third (>4.62s) was flat, likely driven by routine operations. The 4.62s breakpoint's closeness to the 4s neutral point may mean it marks the transition to routine monitoring. These dataset effects may be important for applying MTTC2 in roundabouts. The 1.46s threshold from the interaction subset suggests MTTC2 might work well for real conflicts. The agreement of different methods around 1.1–1.5s may support this. But in mixed traffic (dull data), the segmented regression may provide a more robust approach. Breakpoints at 1.4s and 4.62s may help separate genuine conflicts from normal circulation. This may imply that acceleration-based indicators need geometry sensitive thresholds, especially in roundabouts where acceleration changes are part of safe driving.

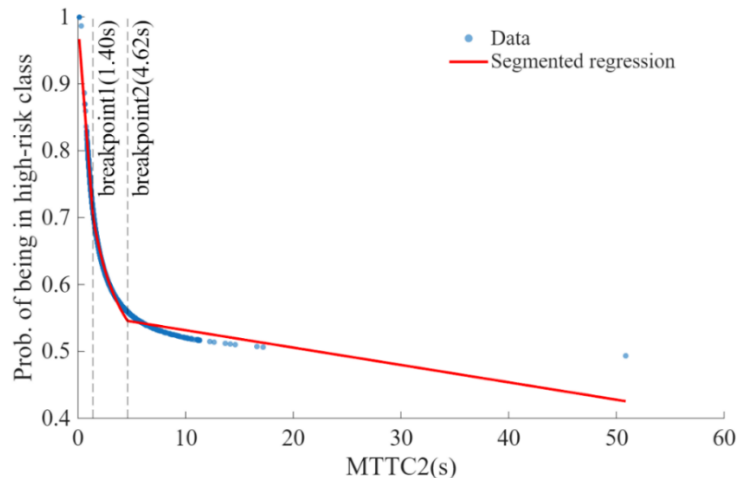

*Figure 5 MTTC2 CBTs based on segmented regression.*

### 3.5. Deeper Insights on Threshold Instability

To move beyond the descriptive findings, we now consider possible reasons for threshold instability and what these might reveal about driver decision-making under varying conditions. When interpreting behavioural thresholds, caution is needed. Threshold stability may depend not only on risk perception but also on how much attention drivers can devote to the task at hand. In our study, the dataset was divided into two parts: an interaction subset, which included only conflict situations where drivers were likely to focus on the immediate risk, and the full dataset, which also included routine navigation where attention was divided across many tasks. This split highlights how threshold patterns may shift depending on whether driving is focused or divided.

We further discuss insights on the difference observed between TTC and MTTC2 threshold stability. It is interesting that particularly in the full dataset, the instability of MTTC2 could point to factors beyond pure risk perception influence the identification of behavioural thresholds. The TTC appeared to maintain interpretable thresholds across both datasets, albeit at different values. The MTTC2 showed substantial instability in the full dataset, with crossover points reaching implausible values (34s). This divergence may not be explained exclusively by differences in the physical formulation of the two indicators, as both are predictive indicators measuring time until collision happens. Instead, it seems plausible that the cognitive demands associated with processing each indicator by a driver could possibly influence whether meaningful thresholds can be identified.

In line with this idea, one tentative hypothesis is that behavioural thresholds can only be observed when drivers are able to cognitively handle and make sense of the indicator being used to measure risk. TTC is based on a simple assumption that vehicles keep moving at their current speeds, something most drivers can quickly judge by asking whether their present speed would lead to a collision. While MTTC2 is more complicated because it requires drivers to consider both speed and acceleration, how these change over time, and how vehicles move along curved paths. Processing all of this information at once may be more than drivers can realistically handle during normal driving.

Patterns in the data may provide some support for this cognitive demand hypothesis. For instance, MTTC2 had a reasonable threshold (1.46s) in the interaction subset where drivers' attention was likely more focused. However, it became unstable in the full dataset where attention was divided across routine navigation tasks. Similarly, the acceleration scale parameter increased noticeably (from $e^{3.075}$ in the full dataset to $e^{1.007}$ in the interaction subset). This may imply that acceleration-related decisions become more consistent when cognitive resources are concentrated on conflict resolution. Another point is the close to zero ASC coefficient estimates in full-dataset models (specially for MTTC2) might not reflect genuinely defensive driving. It could be reflecting cautious behaviour arising from uncertainty when drivers cannot reliably interpret the available information. These observations may help in developing a framework that explains how cognitive load influences both the identification of thresholds and the decision-making strategies drivers apply. Under low cognitive load, when indicators are simple and attention is focused, drivers may engage in utility maximization, optimizing across multiple objectives such as safety, efficiency, and comfort. This could produce the stable thresholds observed for TTC in the interaction subset (around 2s), reflecting genuine risk perception boundaries where drivers make calculated trade-offs.

Under moderate cognitive load, which may result from either greater indicator complexity or divided attention, drivers might adopt a hybrid approach. In this state, their decisions may blend utility maximization with regret minimization, which means that they focus more on avoiding worst-case outcomes than on optimizing overall utility. This shift could explain the greater variability but still interpretable thresholds for TTC in the full dataset (5–7s), where the added caution from regret avoidance creates a wider range of responses. Under high cognitive load, when indicators are complex and attention is divided, drivers may stop trying to process the indicator altogether. Instead, they may fall back on simpler perceptual cues such as distance, visual looming, or rough time-to-contact judgments. The MTTC2 in the full dataset results might reflect this cognitive neglect. The model may be forcing a threshold interpretation on behaviour that is actually driven by entirely different, simpler cues that only weakly correlate with MTTC2 values.

At the same time, other explanations could also be possible. The MTTC2 definition may be less stable in the curved trajectories common at roundabouts. The instability might also reflect real differences in how drivers interpret acceleration. Even so, the observed link between indicator complexity, dataset type, and threshold stability, and the apparent shift from deliberate, multi-objective decision-making toward simpler rule-based strategies, seems to provide at least some tentative support for the role of cognitive limits. If cognitive processing limitations do indeed influence threshold stability and decision strategies, there may be several implications for surrogate safety assessment. First, indicator selection may need to prioritize those that drivers can plausibly process under realistic cognitive loads. Simpler indicators might yield more valid behavioural insights than complex ones, even if the latter are theoretically more precise. Second, thresholds could be context-sensitive, varying not only across driving environments with different cognitive demands but also in their fundamental meaning, with some representing true risk perception points and others merely artifacts of simplified heuristic application. Third, highly unstable thresholds may need careful interpretation. They might indicate cognitive overload and a shift to alternative decision strategies rather than simple measurement error. In such cases, searching for a true threshold could be unsuccessful if drivers are not actually using that indicator in their decision-making process.

### 3.6. Resampling Stability Check

To examine the robustness of the CBTs, we conducted a stability check using ten random samples, each comprising 80% of the utilized data. Tables 6-9 present the model estimation results using the full dataset and the interaction subset of the four CBTs by randomly sampling the data and re-estimating the thresholds using the full data and interaction subset of each conflict indicator. The resampling results provide a first indication of how stable different CBTs may be. For TTC in the full dataset, the crossover point varied between 5.53s and 7.40s across the ten subsamples. At this stage, it is not clear whether such a spread should be viewed as narrow or wide in behavioural terms. A two-second band could reflect genuine heterogeneity in driver responses, but it could also signal sensitivity in the estimation procedure. More systematic checks, for example through bootstrapping or k-fold cross-validation, could help clarify whether the observed variation reflects stable properties of the data or sampling artefacts. Importantly, the threshold did not collapse to implausible extremes as observed for MTTC2, which suggests at least some degree of robustness. In contrast, TTC in the interaction subset had much tighter ranges (1.94–2.11s). This may suggest higher stability when the dataset is restricted to genuine conflict cases. For MTTC2, the contrast was more striking. In the full dataset, the crossover estimates ranged from about 4.7s to over 38s indicating high instability. While in the interaction subset, the range narrowed considerably (1.42–1.57s), aligning with earlier thresholds identified through EVT and segmented regression. These findings may imply that stability itself is dataset-dependent. For full datasets that include many routine, non-conflict events, instability appears more likely. For subsets restricted to genuine conflicts, stability seems much improved. These results tentatively suggest that resampling stability checks could be a useful diagnostic step, but further work is needed to define what should count as acceptable variability and to test robustness across different sampling and modelling approaches.

*Table 6 Stability check of the TTC for CBTs using full data*

| Random Sample (full data) | Mean LL | Inflection Point (s) | Crossover Point (s) | Alertness Point (s) | Neutral Point (s) |
|---|---|---|---|---|---|
| Random Sample 1 | -1.957 | 0.74 | 7.43 | 5.87 | 6.86 |
| Random Sample 2 | -1.956 | 0.74 | 5.62 | 4.67 | 6.95 |
| Random Sample 3 | -1.955 | 0.74 | 6.71 | 5.32 | 6.74 |
| Random Sample 4 | -1.956 | 0.74 | 6.13 | 4.95 | 6.93 |
| Random Sample 5 | -1.956 | 0.74 | 5.97 | 4.72 | 6.99 |
| Random Sample 6 | -1.960 | 0.91 | 5.53 | 4.67 | 6.81 |
| Random Sample 7 | -1.956 | 0.74 | 5.98 | 4.81 | 6.72 |
| Random Sample 8 | -1.955 | 0.74 | 5.62 | 4.56 | 6.74 |
| Random Sample 9 | -1.956 | 0.74 | 5.85 | 4.75 | 6.76 |
| Random Sample 10 | -1.955 | 0.74 | 7.05 | 5.46 | 6.84 |
| [Min, Max.] (Mean) | [-1.960, -1.955] (-1.956) | [0.74, 0.91] (0.76) | [5.53, 7.4] (6.19) | [4.56, 5.87] (4.98) | [6.72, 6.99] (6.83) |
| Sample size | 36,896 | | | | |

*Table 7 Stability check of the TTC for CBTs using interaction data*

| Random Sample (full data) | Mean LL | Inflection Point (s) | Crossover Point (s) | Alertness Point (s) | Neutral Point (s) |
|---|---|---|---|---|---|
| Random Sample 1 | -1.914 | 1.11 | 1.98 | 1.95 | 7.72 |
| Random Sample 2 | -1.915 | 1.14 | 2.05 | 2.00 | 7.64 |
| Random Sample 3 | -1.920 | 1.07 | 2.11 | 2.07 | 7.74 |
| Random Sample 4 | -1.937 | 1.03 | 2.01 | 1.98 | 7.50 |
| Random Sample 5 | -1.919 | 1.06 | 2.09 | 2.06 | 7.74 |
| Random Sample 6 | -1.920 | 1.12 | 1.98 | 1.94 | 7.62 |
| Random Sample 7 | -1.924 | 1.13 | 1.97 | 1.93 | 7.40 |
| Random Sample 8 | -1.907 | 1.15 | 1.99 | 1.96 | 7.59 |
| Random Sample 9 | -1.929 | 1.12 | 1.98 | 1.95 | 7.75 |
| Random Sample 10 | -1.929 | 1.09 | 1.94 | 1.90 | 7.62 |
| [Min, Max.] (Mean) | [-1.937, -1.907] (-1.921) | [1.03, 1.15] (1.10) | [1.94, 2.11] (2.01) | [1.90, 2.17] (1.97) | [7.40, 7.75] (7.63) |
| Sample size | 1,636 | | | | |

*Table 8 Stability check of the MTTC2 for CBTs using full data*

| Random Sample (full data) | Mean LL | Inflection Point (s) | Crossover Point (s) | Alertness Point (s) | Neutral Point (s) |
|---|---|---|---|---|---|
| Random Sample 1 | -1.958 | 0.72 | 34.72 (*1.39**) | 5.81 | 4.73 (*4.70**) |
| Random Sample 2 | -1.957 | 0.72 | 4.89 (*1.37**) | 4.01 | 4.74 (*4.58**) |
| Random Sample 3 | -1.955 | 0.72 | 6.52 (*1.37**) | 4.65 | 4.76 (*4.58**) |
| Random Sample 4 | -1.956 | 0.72 | 31.29 (*1.40**) | 6.07 | 4.80 (*4.70**) |
| Random Sample 5 | -1.957 | 0.72 | 19.69 (*1.38**) | 5.75 | 4.77 (*4.62**) |
| Random Sample 6 | -1.960 | 0.72 | 4.68 (*1.35**) | 3.97 | 4.81 (*4.54**) |
| Random Sample 7 | -1.957 | 0.72 | 6.51 (*1.37**) | 4.64 | 4.76 (*4.54**) |
| Random Sample 8 | -1.956 | 0.79 | 5.30 (*1.37**) | 4.21 | 4.77 (*4.54**) |
| Random Sample 9 | -1.957 | 0.82 | 5.87 (*1.38**) | 4.51 | 4.80 (*4.60**) |
| Random Sample 10 | -1.956 | 0.72 | 38.33 (*1.40**) | 5.75 | 4.72 (*4.68**) |
| [Min, Max.] (Mean) | [-1.960, -1.955] (-1.957) | [0.72, 0.82] (0.73) | [4.68, 38.83] (15.57) *[1.35 1.40] (1.38)* * | [3.97, 6.07] (4.94) | [4.72, 4.81] (4.77) *[4.54, 4.70] (4.61)* * |
| Sample size | 36,896 | | | | |

* *The italicized text indicates the breakpoints obtained from the segmented regression.*

*Table 9 Stability check of the MTTC2 for CBTs using interaction data*

| Random Sample (full data) | Mean LL | Inflection Point (s) | Crossover Point (s) | Alertness Point (s) | Neutral Point (s) |
|---|---|---|---|---|---|
| Random Sample 1 | -1.919 | 0.93 | 1.57 | 1.53 | 4.97 |
| Random Sample 2 | -1.937 | 1.00 | 1.43 | 1.41 | 4.71 |
| Random Sample 3 | -1.928 | 0.97 | 1.49 | 1.46 | 4.85 |
| Random Sample 4 | -1.917 | 0.98 | 1.51 | 1.48 | 4.88 |
| Random Sample 5 | -1.935 | 0.95 | 1.48 | 1.46 | 4.82 |
| Random Sample 6 | -1.939 | 0.90 | 1.48 | 1.46 | 4.81 |
| Random Sample 7 | -1.936 | 0.72 | 1.46 | 1.44 | 4.79 |
| Random Sample 8 | -1.946 | 0.93 | 1.42 | 1.40 | 4.71 |
| Random Sample 9 | -1.948 | 0.90 | 1.45 | 1.42 | 4.79 |
| Random Sample 10 | -1.947 | 0.92 | 1.48 | 1.46 | 4.83 |
| [Min, Max.] (Mean) | [-1.948, -1.917] (-1.935) | [0.72, 1.00] (0.92) | [1.42, 1.57] (1.48) | [1.40, 1.53] (1.45) | [4.71, 4.97] (4.82) |
| Sample size | 1,548 | | | | |

### 3.7. Validation of Candidate Behavioural Thresholds (CBTs)

To evaluate whether a CBT value, such as the crossover point or the inflection point, represents a valid behavioural shift, it is necessary to assess whether incorporating the proposed value improves the model's ability to distinguish between latent behavioural classes. The results in Table 10 provide insight into whether incorporating a proposed threshold may improve the model's ability to capture latent behavioural transitions. In the interaction subset, Models A and B serve as the baseline specifications, where the CM is defined using a continuous TTC transformation (1/TTC) together with an ASC. Models E and F, by contrast, include only the ASC in the CM probability. This setup allows us to examine whether adding the 1/TTC term provides evidence of a behavioural shift beyond the baseline level of caution. The LRTs reported in Table 10 suggest that including the 1/TTC term may lead to a statistically detectable improvement in model fit compared with models that rely only on the ASC. This indicates that TTC could contribute explanatory power beyond baseline caution and may capture a potential shift in behavioural class membership. The results are consistent for both the interaction subset and the full dataset, although the larger magnitude of the LRT in the full dataset may partly reflect the greater number of observations. These outcomes tentatively support the interpretation that the crossover point identified for TTC corresponds to a behaviourally relevant threshold, as drivers may be more likely to transition between low- and high-risk classes in response to changes in TTC rather than being governed solely by baseline caution. Attempts to use the Chow test and LRT in a binary threshold formulation did not produce significant evidence of improvement. This outcome likely reflects data imbalance and the reduced sample size once observations were split at the CBT value, rather than a definitive absence of behavioural change. The continuous specification appears more robust for the available data, while the binary tests may require larger and more balanced samples to provide conclusive results.

*Table 10 Model comparison results for evaluating the behavioural relevance of the TTC crossover point*

| Model | Mean LL | Number of observations | $df$ | AIC | BIC | LRT | $p$ |
|---|---|---|---|---|---|---|---|
| *Interaction Subset* | | | | | | | |
| A | -1.970 | 2,045 | 13 | 4054.691 | 4127.792 | 185.15 | 0 |
| E | -1.925 | | 14 | 3964.114 | 4042.838 | | |
| *Full Data* | | | | | | | |
| B | -1.959 | 46,120 | 13 | 90383.382 | 90496.989 | 289.63 | 0 |
| F | -1.956 | | 14 | 90240.565 | 90362.911 | | |

### 4. Conclusions

This study proposes a behavioural modelling framework to identify thresholds in traffic conflict indicators. The motivation is to complement current statistical approaches, such as EVT, with insights into how drivers may perceive and respond to collision risk. EVT offers rigorous statistical cut-offs based on extremes but does not explicitly have a connection to behavioural processes. A behavioural perspective may provide an additional lens to understand when and how drivers transition between different behavioural states as conflict severity changes.

The proposed framework applies an LC-LK model that captures both inter- and intra-class heterogeneity. This structure distinguishes between low-risk (routine) and high-risk (defensive) driving classes, while allowing correlations among manoeuvres through shared error terms. Conflict indicator probabilities are linked to CM probabilities through a functional transformation, enabling the identification of CBTs from the resulting probability curves. The framework further incorporates diagnostic tools, alternate CBT estimation methods, and validation tests to assess threshold stability. The identification of CBTs follows a systematic process that considers five candidate thresholds: the inflection point (maximum behavioural sensitivity), crossover point (equal probability between classes), alertness point (upper boundary of defensive behaviour), neutral point (where defensive responses become rare), and comfort zone (asymptotic baseline caution). Some of these thresholds may be unreliable in practice, particularly when probability curves are steep or data in the distribution tails is sparse. To address this, the framework suggests diagnostic checks and alternate estimation methods such as segmented regression, the Kneedle algorithm, or change-point detection. Validation tools including Chow tests and likelihood ratio tests are also proposed to assess whether behavioural shifts are statistically meaningful.

The framework is applied to naturalistic roundabout trajectories. For each indicator, the analyses consider Application to naturalistic roundabout trajectories revealed distinct patterns between TTC and MTTC2. For TTC, thresholds remained interpretable across both the full dataset and the interaction subset, though values shifted. The inflection point was concentrated in the 0.8–1.1s range, with crossover points near 2s in the interaction subset and higher (around 6s) in the full dataset, likely influenced by non-conflict cases. By contrast, MTTC2, which incorporates vector-based acceleration, produced lower inflection points (0.6–1.0s) but exhibited substantial instability in the full dataset. Crossover points reached implausibly high values (*e.g.*, 34s) with long flat probability tails, indicating the type of instability highlighted in the diagnostic framework. These findings, together with shifts in ASC coefficients and acceleration parameters, led to the development of a tentative hypothesis. We hypothesize that behavioural thresholds may only emerge when drivers are cognitively able to process the conflict indicator. TTC, with its simple constant-velocity projection, likely falls within drivers' intuitive capacity, while MTTC2, which requires simultaneous processing of acceleration and curved trajectories, may exceed cognitive limits during routine driving. This cognitive load hypothesis offers a tentative explanation for why thresholds appeared stable in interaction only subsets (where attention was focused) but unstable in the full dataset (where attention was divided).

The estimation results also point to broader behavioural distinctions. The baseline propensity for high-risk membership ($\beta_{ASC}$) is strongly negative in interaction-only models but is close to zero in the full dataset. This implies a relatively high baseline caution even in non-conflict conditions. Conflict indicator coefficients ($\beta_{CI}$) are larger in interaction-only models, suggesting heightened responsiveness when drivers are actively engaged in conflicts. Error structure parameters further suggest that acceleration becomes a more standardized evasive response during interactions, but routine driving maybe more variable. Class specific utilities align with these patterns. It shows that low-risk drivers strongly avoid unnecessary deceleration, while high-risk drivers amplify defensive responses in interactions. Stability checks across subsamples indicate that TTC and MTTC2 thresholds are reasonably consistent. Comparison with EVT/GPD thresholds (1.57s for TTC and 1.1s for MTTC2) shows that these fall within the range bounded by the inflection and alertness points. This indicates some alignment between statistical rarity and behavioural sensitivity. However, such alignment may not always hold true. These perspectives appear complementary rather than interchangeable. EVT continues to provide rigorous statistical benchmarks, while the proposed behavioural framework offers interpretive markers that help to contextualise how and when drivers may adjust their responses.

The cognitive load hypothesis remains hypothetical, and further work is required. Future research could develop models that explicitly account for cognitive processing constraints. For example, continuous mixture structures could be used where the probability of processing an indicator varies with its complexity and available cognitive resources. Such models could incorporate perception functions that degrade under cognitive load, explaining threshold instability as a natural consequence of information processing limits rather than measurement error. Empirical validation could employ eye-tracking or neurophysiological sensors in simulators to test whether drivers sample information differently for simple versus complex indicators or vary cognitive load experimentally to assess threshold consistency.

Until such evidence is available, researchers should consider that highly unstable thresholds might imply cognitive overload rather than model inadequacy. This highlights the importance of accounting for human information processing limitations when selecting and interpreting SSMs in practice. Future research could also extend the framework in several directions. Allowing for more than two latent classes may uncover finer distinctions in behaviour, such as aggressive, normal, cautious, or emergency modes. Considering combinations of temporal and spatial indicators would better capture the multi-dimensional nature of risk perception. Most importantly, validation across multiple datasets and traffic environments will be necessary to assess robustness and generalisability. At this stage, the framework relies on SSMs and naturalistic trajectory data. The results suggest that crash data are not strictly required to identify meaningful behavioural thresholds. However, future work could incorporate crash records to validate whether the thresholds observed here align with actual safety outcomes, providing an additional layer of external validity.

**Funding Resource**

The first author is funded by Canda Research Chair Fund. The second author is funded by China Scholarship Council (grant no. 202406560076).